\documentclass[two column]{aastex63}   	
\usepackage{graphicx}				
\usepackage{amssymb, rotating}
\usepackage{caption}

\begin{document}

\title{New Constraints on Protoplanetary Disk Gas Masses in Lupus}

\author[0000-0002-8310-0554]{Dana E. Anderson}
\affiliation{Department of Astronomy, University of Virginia, 530 McCormick Road, Charlottesville, VA 22904, USA}
\affiliation{Virginia Initiative on Cosmic Origins Fellow}

\author[0000-0003-2076-8001]{L. Ilsedore Cleeves}
\affiliation{Department of Astronomy, University of Virginia, 530 McCormick Road, Charlottesville, VA 22904, USA}

\author[0000-0003-0787-1610]{Geoffrey A. Blake}
\affiliation{Division of Geological and Planetary Sciences, California Institute of Technology, 1200 E. California Blvd., Pasadena, CA 91125, USA}

\author[0000-0003-4179-6394]{Edwin A. Bergin}
\affiliation{Department of Astronomy, University of Michigan, 1085 S. University, Ann Arbor, MI 48109, USA}

\author[0000-0002-0661-7517]{Ke Zhang}
\affiliation{Department of Astronomy, University of Wisconsin-Madison, 475 N. Charter Street, Madison, WI 53706, USA}

\author[0000-0003-2251-0602]{John M. Carpenter}
\affiliation{Joint ALMA Observatory, Av. Alonso de C\'{o}rdova 3107, Vitacura, Santiago, Chile}

\author[0000-0002-6429-9457]{Kamber R. Schwarz}
\affiliation{Max-Planck-Institut für Astronomie, Königstuhl 17, 69117 Heidelberg, Germany}

\begin{abstract} \noindent Gas mass is a fundamental quantity of protoplanetary disks that directly relates to their ability to form planets. Because we are unable to observe the bulk H$_2$ content of disks directly, we rely on indirect tracers to provide quantitative mass estimates. Current estimates for the gas masses of the observed disk population in the Lupus star-forming region are based on measurements of isotopologues of CO. However, without additional constraints, the degeneracy between H$_2$ mass and the elemental composition of the gas leads to large uncertainties in such estimates. Here we explore the gas compositions of seven disks from the Lupus sample representing a range of CO-to-dust ratios. With Band 6 and 7 ALMA observations, we measure line emission for HCO$^+$, HCN, and N$_2$H$^+$. We find a tentative correlation among the line fluxes for these three molecular species across the sample, but no correlation with $^{13}$CO or sub-mm continuum fluxes. For the three disks where N$_2$H$^+$ is detected, we find that a combination of high disk gas masses and sub-interstellar C/H and O/H are needed to reproduce the observed values. We find increases of $\sim$10--100$\times$ previous mass estimates are required to match the observed line fluxes. This study highlights how multi-molecular studies are essential for constraining the physical and chemical properties of the gas in populations of protoplanetary disks and that CO isotopologues alone are not sufficient for determining the mass of many observed disks. \end{abstract}

\section{Introduction}
The amount of material present in protoplanetary disks determines their continued potential for planet formation.  The dominant contributor to the disk mass is molecular hydrogen, H$_2$. As a homonuclear diatomic molecule, H$_2$ lacks a dipole moment and only produces rovibrational bands and pure rotational lines through weak quadruple transitions. In addition, the spacing of the fundamental ground state energy levels of H$_2$ further inhibits emission from cold ($\sim$20~K) gas. All together, this renders the bulk H$_2$ gas in disks observationally untraceable. We therefore rely on indirect tracers combined with empirical conversion factors or astrochemical modeling to determine total disk masses \citep[see review by][]{2017ASSL..445....1B}. The most common molecular gas mass tracers are the isotopologues of CO. CO is abundant in disks and its rotational emission is readily detectable at submm/mm wavelengths \citep[e.g.,][]{2016ApJ...828...46A, 2016ApJ...827..142B, 2017ApJ...844...99L, 2019ApJ...872..158A}. Given the narrow range in $^{12}$CO/H$_2$ of $\sim$ 0.5--3.0$\times$10$^{-4}$, CO is found to be a reliable tracer of H$_2$ in interstellar environments until freeze-out becomes important under pre-stellar conditions \citep{2017ASSL..445....1B}. But, as we discuss below, there is currently a great deal of uncertainty regarding the CO-to-H$_2$ conversion factor in protoplanetary disks. 

\begin{deluxetable*}{llcccccccccccc}
\tablewidth{0pt}
\setlength{\tabcolsep}{4pt}
\tablecolumns{12}
\tablecaption{\em{Source Properties and Line Fluxes}\label{table1}}
\tablehead{\multicolumn{1}{l}{ID}&\multicolumn{1}{l}{Source} & \multicolumn{1}{c}{SpT} & \multicolumn{1}{c}{dist.} & \multicolumn{1}{c}{ $L_*$} & \multicolumn{1}{c}{$M_*$} & \multicolumn{1}{c}{ log~$\dot{M}\mathrm{_{acc}}$} & \multicolumn{1}{c}{$M\mathrm{_{dust}}$}  & \multicolumn{5}{c}{Line Fluxes  (mJy km s$^{-1}$), $J$~=~3--2}  \\
 \# & Name  & & (pc) & ($L_\odot$) &($M_\odot$) & ($M_\odot$ yr$^{-1}$) & ($M_\oplus$)   & $^{13}$CO & C$^{18}$O & HCO$^+$ & N$_2$H$^+$ & HCN \\
 & & [1] & [1] & [1] &  [1] & [1] & [2] & [2] & [2] & [3] & [3] & [3] }
\startdata
1 & Sz 65  & K7 & 155 &  0.89 & 0.75 & -9.55 & 15.16 $\pm$ 0.08  & 971 $\pm$ 128& 415 $\pm$ 105 & 219 $\pm$ 23 & $<$19 &  $<$32  \\
2 & Sz 130& M2 & 160 & 0.18 & 0.37 & -9.06 & ~1.45 $\pm$ 0.08 &  470 $\pm$ ~71 & $<$117 & 184 $\pm$ 20 & $<$27 & $<$37  \\
3 & Sz 68 & K2 & 154 & 5.42 & 2.15 & -8.39 & 35.34 $\pm$ 0.11  &  915 $\pm$ 133 &	444 $\pm$ 132 & 183 $\pm$ 20 & $<$15	& $<$24 \\
4 & Sz 90 & K7 & 160 & 0.42 & 0.80  & -8.97 & ~9.12 $\pm$ 0.19  & 	433 $\pm$ ~98 &	$<$183 & 298 $\pm$ 31 & 70 $\pm$ 10 & 169 $\pm$ 26  \\
5 & J16085. & M3 & 168 & 0.21  & 0.31 & -10.02 & ~8.17 $\pm$ 0.12  & 267 $\pm$ ~46 & $<$135 & 254 $\pm$ 26 & $<$36 & 162 $\pm$ 19 \\
6 & Sz 118 & K5 & 164 & 0.72 & 1.10 & -9.24 & 26.41 $\pm$ 0.41 &	695 $\pm$ 133	& $<$234 & 376 $\pm$ 38 & 64 $\pm$ 11 & 175 $\pm$ 21 \\
7 & Sz 129 & K7 & 162 & 0.43 & 0.82 & -8.33 & 42.56 $\pm$ 0.12  & 516 $\pm$ ~72	& $<$108 & 595 $\pm$ 61 & 203 $\pm$ 27 & 443 $\pm$ 46 
\enddata
\tablecomments{Sz~68 is a triple stellar system \citep{2018ApJ...869L..44K}. The ID number listed in the first column is used to identify the sources in Figures~\ref{fig:obsFlux}--\ref{fig:stats}.}
\tablereferences{[1] \citet{2019AA...629A.108A}, [2] \citet{2016ApJ...828...46A}, [3] This work, as described in Section~\ref{section:methods2}.}
\end{deluxetable*}

The Atacama Large Millimeter/submillimeter Array (ALMA) disk survey by \citet{2016ApJ...828...46A} uses observations of $^{13}$CO and C$^{18}$O to explore the gas masses of disks in the Lupus star-forming region (1--3~Myr in age, at a distance of 150--200~pc). \citet{2016ApJ...828...46A} studied a total of 89 disks. Within this sample, 62 were detected in continuum, 36 in $^{13}$CO and 11 in C$^{18}$O. They found typical gas masses of $\lesssim$1 Jupiter mass based on the CO isotopologue fluxes and assuming an interstellar CO/H$_2$ abundance. \cite{2017AA...599A.113M} produced disk gas mass estimates for the Lupus disks with $^{13}$CO detections using a large grid of $\sim$800 physical-chemical disk models. They simulated continuum and CO isotopologue fluxes for a range of stellar and disk properties, including disk masses from 10$^{-5}$--10$^{-1}$~M$_\odot$. Even when accounting for freeze-out and isotopic processes, most of the Lupus disks appear to have gas-to-dust mass ratios of $\sim$1--10 when interstellar volatile carbon and oxygen abundances are assumed. Such gas masses would suggest that substantial H$_2$ mass loss has occurred relative to the assumed initial interstellar gas-to-dust ratio of 100. Given the 1--3~Myr age of the Lupus sources, these results present a limited time window for the formation of gas giant planets in these disks. A similar survey of the Chamaeleon~I region also found faint CO isotopologue emission, indicating that this may be a prevalent phenomenon in Class~II disks \citep{2017ApJ...844...99L}.    

This strict time constraint could be alleviated if the $^{13}$CO and C$^{18}$O fluxes are, in fact, not representative of the total H$_2$ mass in disks. An alternative hypothesis is that the gas in the outer regions of the aforementioned disks surveyed by ALMA is depleted in carbon and/or oxygen relative to interstellar abundances \citep{2016ApJ...828...46A,2017AA...599A.113M,2017ApJ...844...99L}. Using a sub-interstellar CO/H$_2$ conversion factor would result in higher H$_2$ gas mass estimates. Detailed analysis of a small number of disks with HD detections from the Herschel Space Observatory provide an independent constraint on the H$_2$ mass. Comparisons of CO isotopologue and HD measurements indicate that CO abundances in protoplanetary disks can be up to two orders of magnitude below the interstellar value \citep{2013Natur.493..644B,2013ApJ...776L..38F,2016ApJ...831..167M,2017A&A...605A..69T,2021arXiv210906228S, 2021ApJ...908....8C}. When comparing different potential H$_2$ gas tracers in the Lupus disks, \cite{2016AA...591L...3M} found that disk gas mass estimates based on sub-mm continuum show a correlation with mass accretion rates, as anticipated by viscous accretion theory. Meanwhile they find that disk gas mass estimates based on the CO isotopologue fluxes show no such correlation suggesting that CO-based disk masses may be lower limits. Yet physical-chemical disk models are unable to determine if the Lupus disks have sub-interstellar CO/H$_2$ abundances (i.e., CO depletion) based on measurements of CO isotopologues and sub-mm continuum alone \citep{2017AA...599A.113M}. We aim to further constrain the gas masses of a subset of the Lupus disks using observations of N$_2$H$^{+}$.

N$_2$H$^{+}$ is formed mainly through the protonation of N$_2$ via H$_3^+$, which is prevalent in ionized H$_2$-rich gas. N$_2$ is expected to be a major nitrogen carrier throughout most of the disk \citep{2014ApJ...797..113S} and has similar volatility to CO \citep{2006AA...449.1297B}. In addition to electron recombination, CO represents one of the main destruction pathways for N$_2$H$^{+}$. CO has a higher proton affinity than N$_2$ resulting in destruction of N$_2$H$^{+}$ to form HCO$^+$ when CO is abundant. Therefore, CO depletion increases not only the gas-phase N/C and N/O elemental ratios, but N$_2$H$^+$ abundances as well. As such, we anticipate N$_2$H$^{+}$ to be an indicator of CO depletion. N$_2$H$^{+}$ has been used in prior observational studies as a tracer of the CO snowline \citep{2013Sci...341..630Q,2017AA...599A.101V} and to infer depletion of gas-phase CO in disks in the Upper Scorpius star-forming region \citep{2019ApJ...881..127A}. To supplement the N$_2$H$^+$ observations, we also include HCO$^+$ and HCN to provide additional constraints on the ionization and elemental composition of the gas.  

We present a chemical survey of seven disks in the Lupus star forming region with gas masses of 10$^{-5}$--10$^{-3}$~M$_\odot$ based on current $^{13}$CO and C$^{18}$O data \citep{2017AA...599A.113M}. The observations and modeling procedure is described in Section~\ref{section:methods}. Comparisons between the measured spectral line fluxes and our disk models is presented in Section~\ref{section:results}. We then discuss these results in Section~\ref{section:discussion} and present our conclusions in Section~\ref{section:conc}.

\begin{figure*}
\includegraphics[width=\linewidth]{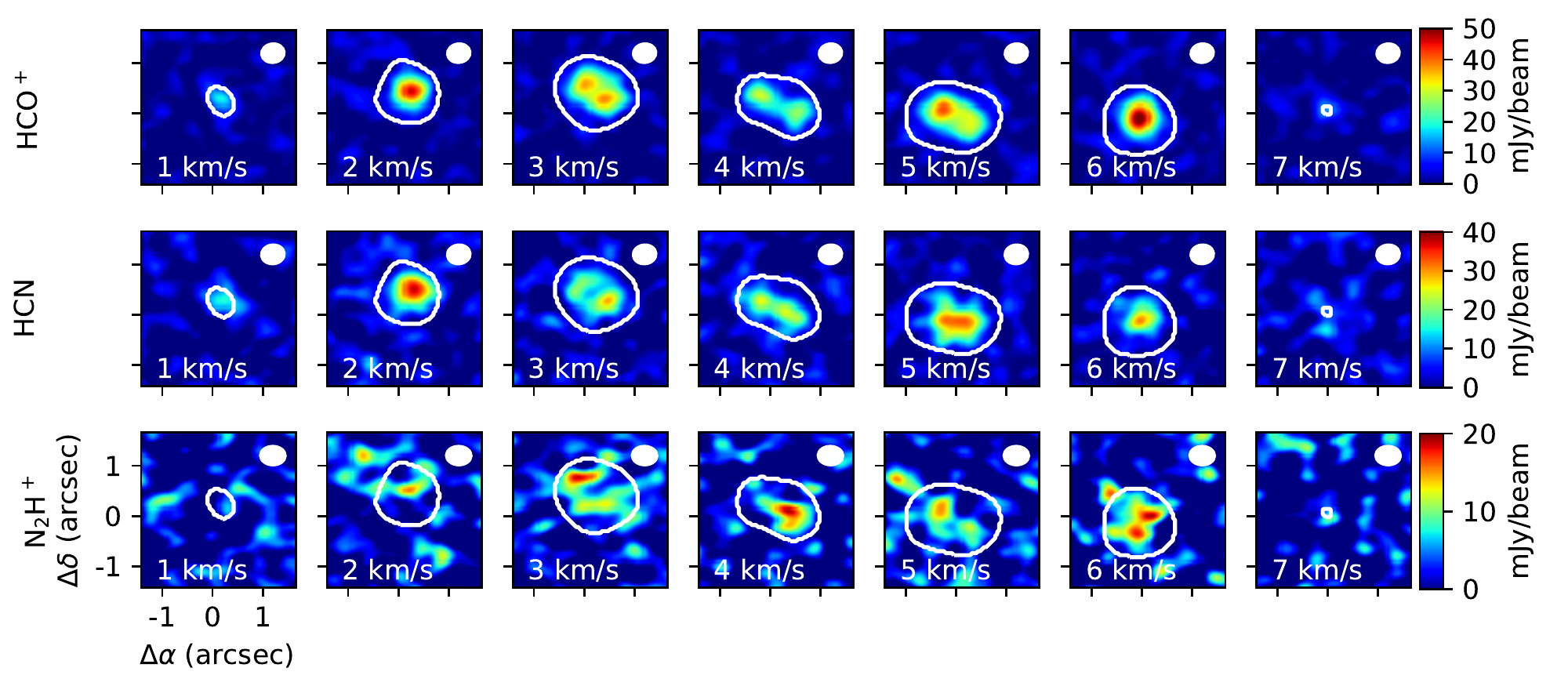}
\caption{Channel maps of HCO$^+$, HCN, and N$_2$H$^+$ line emission for Sz~129. Contours outline the mask generated based on the HCO$^+$ emission. Tick marks indicate 1\arcsec. The beam is shown in the upper right. Channel maps for all seven sources are provided in the appendix. \label{fig:maps7}}
\end{figure*}

\section{Methods}\label{section:methods}
To investigate the gas properties of our sample of Lupus disks, we used a combination of ALMA observations and physical-chemical modeling. The selected disks have previously measured $^{13}$CO~$J$~=~3--2 and sub-mm continuum fluxes \citep{2016ApJ...828...46A}. Our goal is to provide further constraints on the disk mass and composition through the observation of additional chemical tracers, namely N$_2$H$^+$, HCO$^+$, and HCN. 

\subsection{Observations}\label{section:methods1}
We searched for N$_2$H$^+$, HCO$^+$, and HCN line emission in a set of T~Tauri disks using ALMA. The sources investigated in this work were taken from the Lupus disk survey by \citet{2016ApJ...828...46A}. Their survey selected 93 sources that had stellar masses $>$0.1~M$_\odot$ and were classified as protoplanetary disks based on having Class~II or flat IR spectra. Of the 93 sources, 89 were observed by ALMA. We chose our sample of 7 sources from the 36 that were detected in both continuum and $^{13}$CO $J$=3--2 emission. Our two brightest disks in $^{13}$CO $J$=3--2 were among the 11 sources also detected in C$^{18}$O $J$=3--2 emission. Seven disks (Sz~65, Sz~130, Sz~68, Sz~90, 2MASSJ16085324-3914401, Sz~118, and Sz~129) were selected from the ALMA survey of the Lupus star-forming region by \cite{2016ApJ...828...46A} to represent a range of observed CO isotopologue fluxes relative to their dust continuum emission. The disks represent a range in estimated gas-to-dust mass ratios from $\sim$0.5 to 30 based on the CO and dust observations \citep{2017AA...599A.113M}. 

The chosen sources surround K- and M-type stars with a range of stellar masses \citep[0.31--2.15~M$_\odot$;][]{2019AA...629A.108A}, stellar luminosities \citep[0.18--5.42~L$_\odot$;][]{2019AA...629A.108A}, and disk dust masses derived from continuum emission \citep[1--40~M$_\oplus$;][]{2016ApJ...828...46A}. Source properties are provided in Table~\ref{table1}. Source distances range from 154 to 168~pc \citep{2019AA...629A.108A}. Sz~68 (HT~Lup) and Sz~129 were included in the DSHARP sample \citep{2018ApJ...869L..41A} and observed at high spatial resolution (35~mas or 5~au) by ALMA. The DSHARP program focused on large, bright disks at relatively close distances ($\sim$100--170~pc). Sz~68 is a triple system in which spiral arms were found in the primary disk \citep{2018ApJ...869L..44K}. For Sz~129, DSHARP confirmed previous tentative detections of annular substructure \citep{2018ApJ...869L..42H}.  

The ALMA Cycle 7 observations (PI: Anderson, 2019.1.01135.S) took place in October and November of 2019. Baselines ranged from 15.1 to 783.5~m with 43--47 antennas in the 12~m array. Two spectral setting were observed. In Band 6, 117.1875~MHz-wide spectral windows were centered at  267.540~GHz (capturing the HCO$^+$ $J$~=~3--2 line), 265.869~GHz (capturing the HCN $J$~=~3--2 line), 255.462~GHz, and 251.783~GHz. A 2~GHz-wide spectral window centered at 269.489~GHz was included to capture continuum emission. In Band~7, 117.1875~MHz-wide spectral windows were centered at  279.495--279.507~GHz (capturing the N$_2$H$^+$ $J$~=~3--2 line), 288.127--288.139~GHz, 289.192--289.204~GHz, and 289.628--289.640~GHz. Two 2~GHz-wide spectral windows, one centered at 278.168--278.180~GHz and one at 290.898--290.910~GHz, were included to capture continuum emission. Observed channel widths for spectral line windows were 122.070~kHz or about 0.14~km~s$^{-1}$ (61.035~kHz or about 0.07~km~s$^{-1}$ in the case of the 251.783~GHz window) in Band~6 and 244.141~kHz or about 0.25~km~s$^{-1}$ (61.035~kHz or about 0.07~km~s$^{-1}$ in the case of the N$_2$H$^+$ line window) in Band~7. The channel width used for continuum windows was 15625~kHz. The on-source integration time was about 0.25~hr per source for Band 6 and 0.5~hr per source for Band~7.

We analyzed the observations using the Common Astronomy Software Applications (CASA) package \citep{2007ASPC..376..127M} version 5.6.1--8. The data were calibrated using the standard ALMA pipeline. Sources J1924-2914, J1427-4206, and J1517-2422 were used for flux and bandpass calibrations in different execution blocks. J1610-3958 was used as the phase calibrator. Self-calibration was attempted, but did not improve the RMS noise level. 

\begin{figure}
\includegraphics[width=\linewidth]{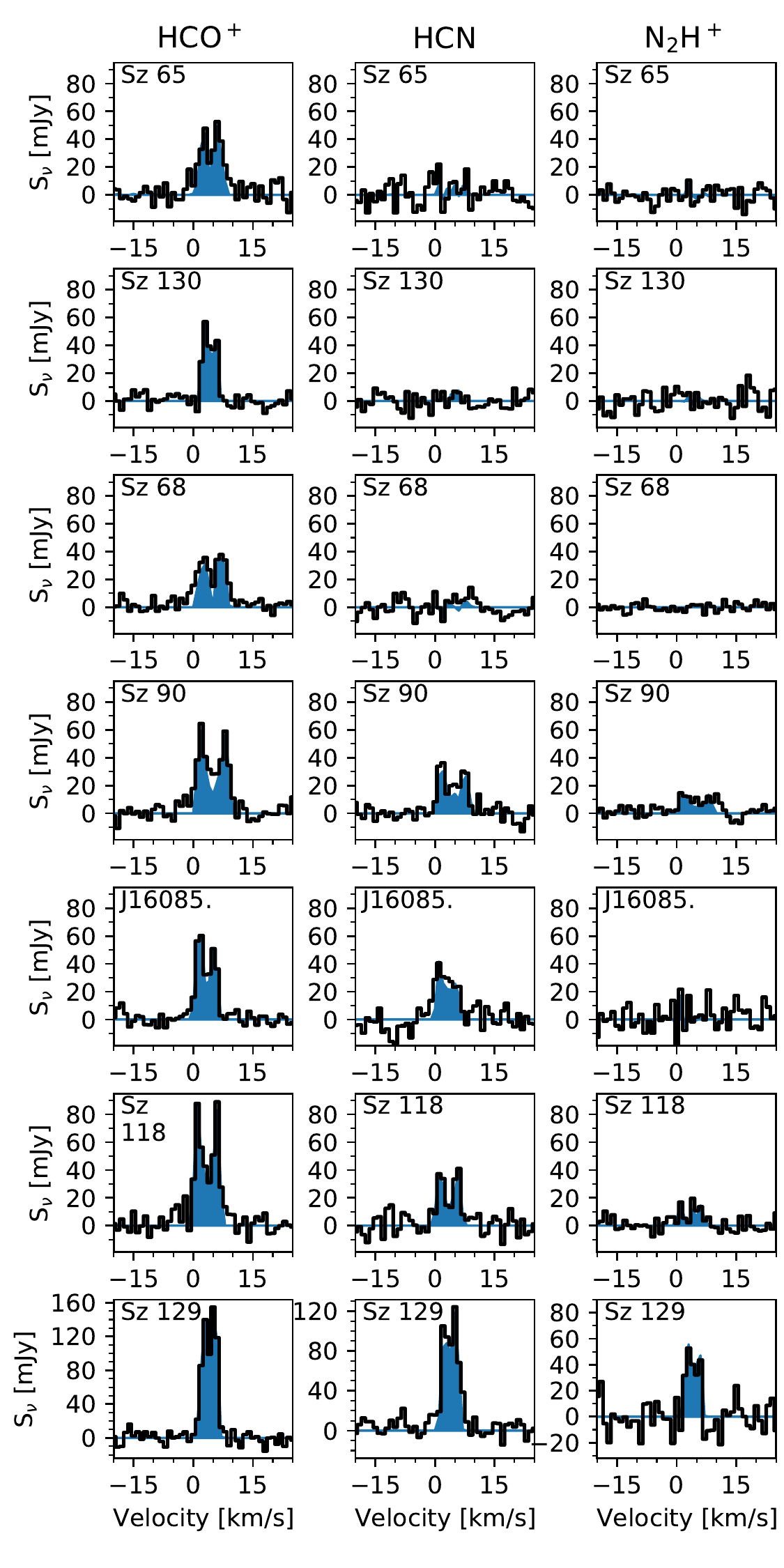}
\caption{HCO$^+$, HCN, and N$_2$H$^+$ $J$~=~3--2 spectra for the observed sources. Emission within the line mask (described in Section~\ref{section:methods2}) is shown in blue. \label{fig:spectra}}
\end{figure}

\subsection{Spectral Line Analysis}\label{section:methods2}
To analyze the line emission, we used CASA's $uvcontsub$ to remove continuum emission from spectral windows. We then imaged the bright HCO$^+$ emission to generate an empirical mask to use for identifying emission from weaker lines, including N$_2$H$^+$. By using a strong line to create an empirical mask, we do not require any physical assumptions such as stellar mass or disk inclination, like are necessary in the generation of Keplerian masks. Using $tclean$ with 1~km~s$^{-1}$ channels and Briggs weighting with a robust parameter of 0.5, we initially selected the regions of HCO$^+$ emission by hand. Smoothing the HCO$^+$ image with a synthetic beam (from about 0.45--0.5\arcsec~$\times$~0.35--0.4\arcsec~to 0.6\arcsec~$\times$~0.6\arcsec) and clipping out emission below the level of the off-source noise, we generated an empirical mask to select the regions in each channel that contained HCO$^+$ emission. This mask was then used as the ``clean mask" in $tclean$ to generate the HCO$^+$, N$_2$H$^+$, and HCN images. For comparison, we also generated images using the automasking algorithm ``auto-multithresh" in $tclean$. We started with the recommended setup for long baselines, then tested noise thresholds of 3--5$\sigma$ combined with minimum beam fractions of 0.5--0.3. We found no noticeable differences in the resulting channel maps compared to those generated with the empirical mask. An example of the channel maps for the targeted HCO$^+$, N$_2$H$^+$, and HCN lines in Sz~129 is shown in Figure~\ref{fig:maps7}. Channel maps for all the sources are provided in Figures~\ref{fig:maps1}--\ref{fig:maps7A}. White contours outline the masked regions based on the observed HCO$^+$ emission.

The empirical masks were also used to generate spectra and the integrated line fluxes. Spectra are shown in Figure~\ref{fig:spectra}. Line emission within the mask is indicated in blue. For comparison, the mask components from each channel were combined over the entire spectral range to create a single mask. Spectra generated using this mask are indicated by the black curve. Table~\ref{table1} provides the calculated line fluxes. After multiplying the per-channel mask by the original image with $immath$, we used the $specflux$ command to calculate the integrated line flux for each of the spectral lines of interest. The corresponding noise estimates were computed by applying the same mask to 30 sets of line-free channels in the image cube and taking the standard deviation of the integrated fluxes. Upper limits are estimated as three times this noise value. The reported noise estimates and upper limits also include an assumed 10\% 1$\sigma$ error on the line flux due to flux calibration added in quadrature. 

\begin{deluxetable}{ll}[b]
\tablewidth{0pt}
\tablecolumns{2}
\tablecaption{\em{Key Disk Model Parameters}\label{table2}}
\tablehead{\colhead{Parameter} & \colhead{Value(s)}}
\startdata
Stellar X-ray Luminosity & 10$^{29}$, 10$^{30}$, 10$^{31}$~erg~s$^{-1}$ \\
Disk Mass & 10$^{-4}$, 10$^{-3}$, 10$^{-2}$~M$_\odot$ \\
C/H, O/H & $\sim$10$^{-6}$, $\sim$10$^{-5}$, $\sim$10$^{-4}$
\enddata
\end{deluxetable}

\subsection{Modeling}\label{section:methods3}
We compare our observed fluxes to predictions from our physical-chemical disk model \citep{2011ApJ...726...29F,2013ApJ...772....5C,2015ApJ...799..204C,2018ApJ...865..155C,2021ApJ...909...55A}. The model framework begins with radiative transfer modeling using TORUS \citep{2004MNRAS.350..565H} to establish the dust temperature for the assumed gas and dust structures. Attenuation of UV and X-ray irradiation by the disk is then computed using the method of \citet{2011ApJ...740....7B,2011ApJ...739...78B} for a T~Tauri star. Gas temperatures are estimated based on the local UV flux and the gas density using fitting functions to the thermochemical models of \cite{2013A&A...559A..46B}, as described in \cite{2015ApJ...799..204C}. The incident cosmic-ray ionization rate at the disk surface is 1.1$\times$10$^{-18}$ s$^{-1}$ and is modulated with vertical depth in the disk \citep[see][]{2015ApJ...799..204C}. 

Using this setup, we generated a series of physical-chemical models for various disk gas masses and initial elemental C/H and O/H abundances. The modeled disks are azimuthally symmetric, encircling a 1~M$_{\odot}$ T~Tauri star with a radius of 2.8~R$_{\odot}$ and an effective temperature of 4300~K. Table~\ref{table2} lists the varied model parameters. The disk structure is shown in the appendix in Figure~\ref{fig:env}. The gas and dust structures are identical to those of \citet{2021ApJ...909...55A}. The dust mass is held at a constant value of 10$^{-4}$~M$_\odot$ across all the modeled disks, while the gas mass is varied from 10$^{-4}$--10$^{-2}$~M$_\odot$ to sample different gas-to-dust mass ratios. Chemical abundances are computed by solving the rate equations for our chemical reaction network (\citealt{2011ApJ...726...29F}; updated in \citealt{2021ApJ...909...55A}) and assumed initial abundances \citep[see][]{2021ApJ...909...55A} for disk radii out to a distance of 100~au from the central star. C/H and O/H abundances were varied together to produce three sets of initial compositions: (1) C/H~=~1.3$\times$10$^{-4}$ and O/H~=~2.1$\times$10$^{-4}$, (2) C/H~=~1.3$\times$10$^{-5}$ and O/H~=~2.1$\times$10$^{-5}$, and (3) C/H~=~1.3$\times$10$^{-6}$ and O/H~=~2.1$\times$10$^{-6}$. The relative amounts of the major initial molecular carbon and oxygen carriers were held constant as the C/H and O/H abundances were varied. The effects of varying the C/O ratio from 0.6 to 1.5 are briefly discussed in Section~\ref{section:discussion1}.

Spectral line fluxes were computed using the non-LTE excitation and radiative transfer code LIME \citep{2010AA...523A..25B} using our modeled chemical abundances at a time of approximately 1~Myr as input. The choice of time may affect the chemical abundances. Abundances of N$_2$H$^+$, HCO$^+$, HCN, and CO vary by less than a factor of 2--3$\times$ for disk radii greater than 20~au from the central star in our fiducial model (disk mass of 10$^{-2}$~M$_\odot$, C/H~=~1.3$\times$10$^{-4}$ and O/H~=~2.1$\times$10$^{-4}$, L$_{X-ray}$~=~10$^{30}$~erg~s$^{-1}$). But for inner radii or disks depleted in total gas or volatiles, chemical abundances are more variable across this time period. $^{13}$CO fluxes were computed using an assumed $^{12}$C/$^{13}$C of 68 \citep{2005ApJ...634.1126M}. Collisional rates were taken from the LAMDA database \citep{2005AA...432..369S}. We use the collision rates without hyperfine splitting for the line emission predictions. The models assume a distance of 160~pc and an inclination of 60$\degr$. 

\begin{figure*}
\includegraphics[width=\linewidth]{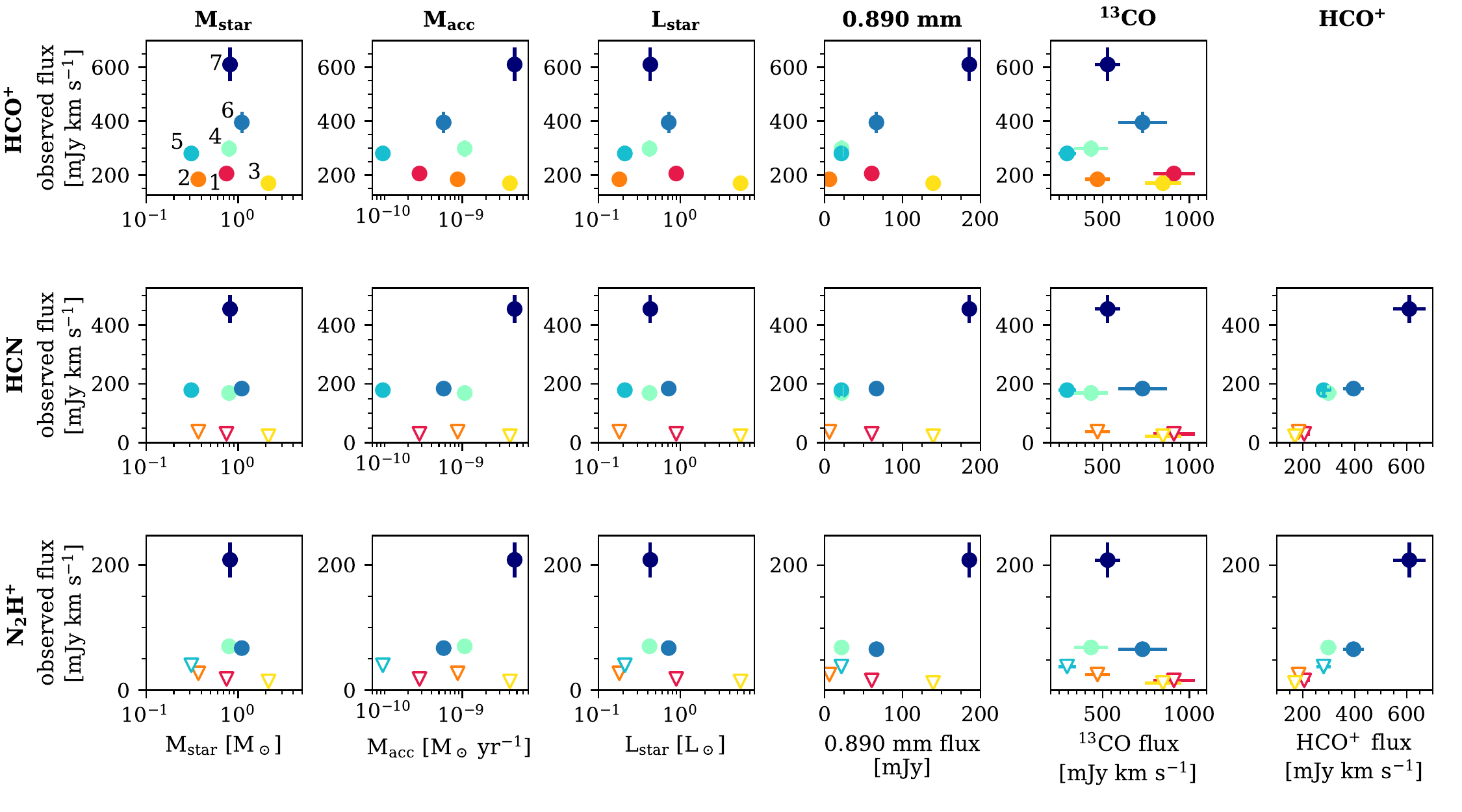}
\caption{Comparison of observed line fluxes for HCO$^+$, HCN, and N$_2$H$^+$ (from top to bottom rows) to stellar mass \cite[first column;][]{2019AA...629A.108A}, mass accretion rate \cite[second column;][]{2019AA...629A.108A}, stellar luminosity \cite[third column;][]{2019AA...629A.108A}, 0.890-mm continuum \cite[fourth column;][]{2016ApJ...828...46A}, $^{13}$CO \cite[fifth column;][]{2016ApJ...828...46A}, and HCO$^{+}$ fluxes (final column; this work). All line fluxes are scaled to a distance of 160~pc; fluxes are multiplied by $\frac{d^2}{(160~pc)^2}$ where $d$ is the source distance in parsecs from Table~\ref{table1}.  Downward triangles represent upper limits. In the first panel, the sources are labeled with ID numbers from Table~\ref{table1}. Colors are used to distinguish the seven different sources.}
\label{fig:obsFlux}
\end{figure*}

\begin{figure}
\includegraphics[width=\linewidth]{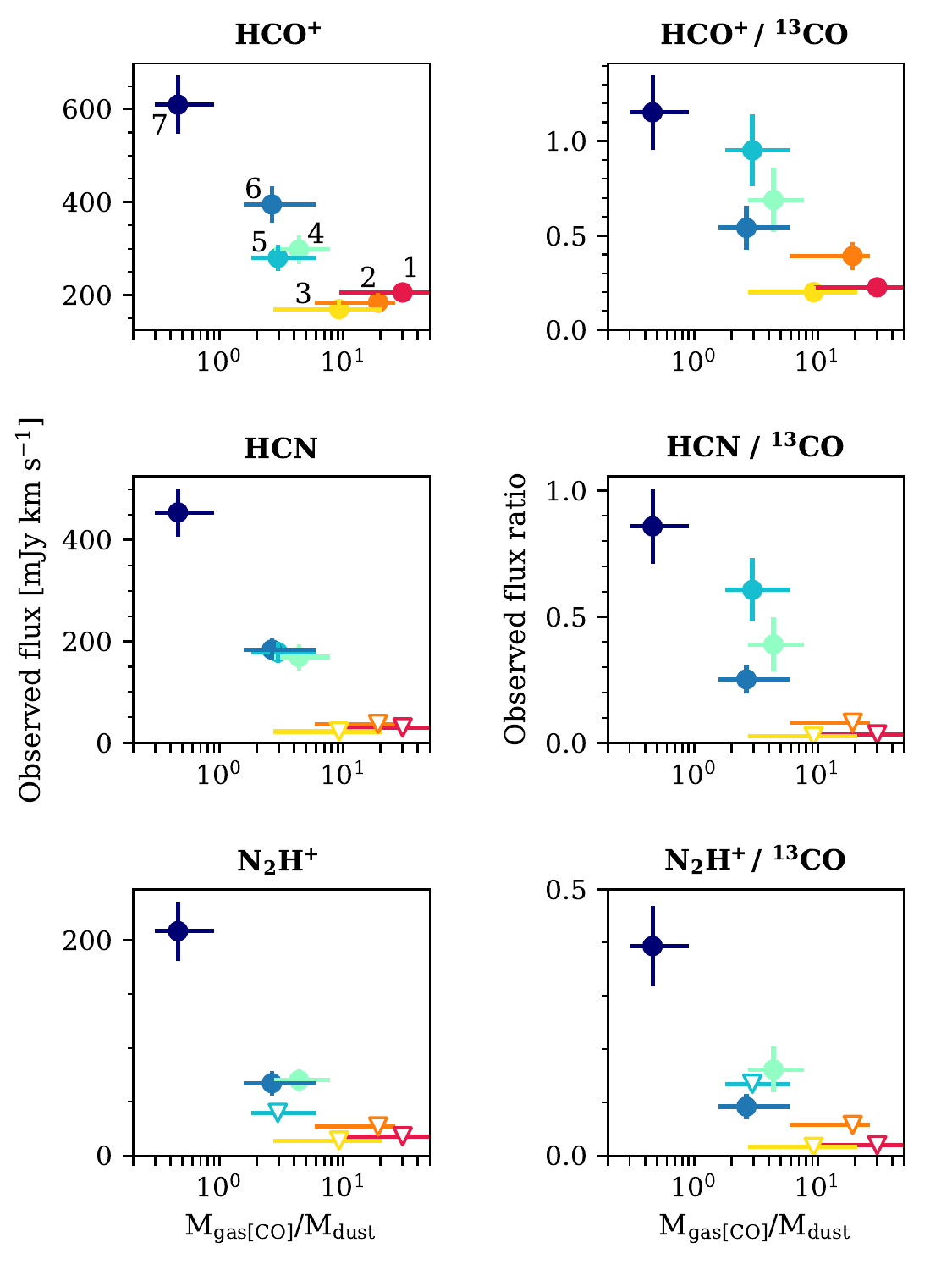}
\caption{Observed HCO$^+$, HCN, and N$_2$H$^+$ fluxes and those relative to $^{13}$CO vs.~the gas-to-dust mass ratios estimated from CO isotopologue fluxes by \cite{2017AA...599A.113M}. All line fluxes are scaled to a distance of 160~pc. Downward triangles represent upper limits. In the first panel, the sources are labeled with ID numbers from Table~\ref{table1}. Colors are used to distinguish the seven different sources. \label{fig:obsFluxRatio}}
\end{figure}

\begin{figure*}
\includegraphics[width=\linewidth]{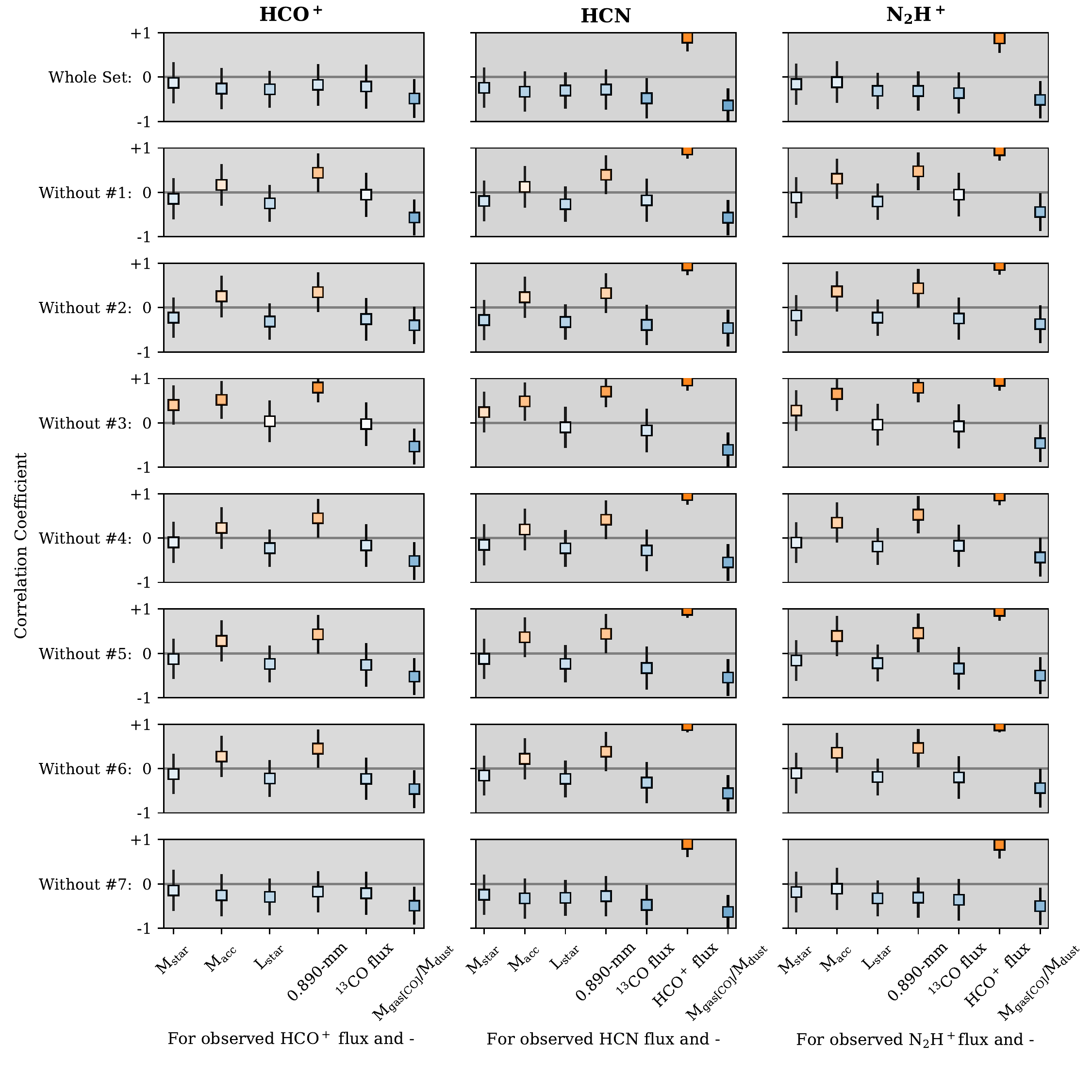}
\caption{Median linear correlation coefficients for each source parameter with observed HCO$^+$, N$_2$H$^+$, and HCN fluxes. The first row corresponds to the calculation for the entire set of seven sources and subsequent rows test for robustness of these results by leaving one source out and calculating the correlation coefficient using data from the remaining six. Looking down each column, correlation coefficients that do not vary when individual disks are removed are more representative of the entire sample rather than driven by individual (possibly outlier) disks. Positive correlations (with coefficients from 0 to 1) are shown in orange and anticorrelations (0 to -1) in blue. Darker colors represent stronger (anti)correlations.}
\label{fig:stats}
\end{figure*}

\section{Results}\label{section:results}
Here we report on the outcome of our spectral line analysis as described in Section~\ref{section:methods1}--\ref{section:methods2}. After discussing the spectral line fluxes for HCO$^+$, N$_2$H$^+$, and HCN, we plot the measured values against available source properties and $^{13}$CO fluxes from the literature. Finally, we compare the observed values with model predictions for various disk gas masses and CO abundances using the model setup from Section~\ref{section:methods3}. 

\subsection{Spectral Line Fluxes}
Based on our spectral line analysis, we find HCO$^+$ emission present in all seven sources with fluxes ranging from 183$\pm$20 to 595$\pm$61~mJy~km~s$^{-1}$. In comparison, N$_2$H$^+$ and HCN fluxes are weaker and only detected in about half of the sample (see Figure~\ref{fig:spectra} and Table~\ref{table1}). For sources with detectable levels of N$_2$H$^+$ and HCN, this emission spans a spectral range and spatial extent similar to that of the HCO$^+$ emission as demonstrated by the channel maps in Figures~\ref{fig:maps7} and \ref{fig:maps1}--\ref{fig:maps7A}. These similarities support the use of the empirical mask for selecting regions of line emission in this analysis.

\subsection{Comparison with Source Properties}
After computing the spectral line fluxes, we compared them to multiple source parameters including the stellar mass, mass accretion rate, stellar luminosity, and previously reported 0.890-mm continuum and $^{13}$CO~$J$~=~3--2 fluxes (see Figure~\ref{fig:obsFlux}). We find no correlation with stellar mass or luminosity or mass accretion rates \cite[from][]{2019AA...629A.108A} for our measured HCO$^+$, N$_2$H$^+$, and HCN fluxes. Nor does the comparison with sub-mm continuum and CO isotopologue fluxes \cite[from][]{2016ApJ...828...46A} reveal any clear trends. Three disks have HCO$^+$ fluxes of $\sim$200~mJy~km~s$^{-1}$, but display a wide range in 0.890-mm fluxes. The spread in dust masses for these three disks extends from $\sim$1 to 35~M$_\oplus$ (relative to the range of $\sim$1 to 42~M$_\oplus$ for the entire sample). The remaining disks have higher HCO$^+$ fluxes that increase with the 0.890-mm flux. There is no obvious trend in N$_2$H$^+$ nor HCN with 0.890-mm flux. The comparison with mass accretion rates looks similar to that with the 0.890-mm flux, with the exception of Sz~90 (labeled as source~4 in Fig.~\ref{fig:obsFlux}). This is not surprising given the correlation between mass accretion rates and dust masses seen in Lupus \citep{2016AA...591L...3M}.  

Sz~129 (labeled as source~7 in Fig.~\ref{fig:obsFlux}) stands out from the rest of the sample with the highest 0.890-mm, HCO$^+$, N$_2$H$^+$, and HCN flux. At the same time, Sz~129 has the median $^{13}$CO flux for this set of seven sources and the lowest of the upper limits on C$^{18}$O. Thus, the HCO$^+$, N$_2$H$^+$, and HCN fluxes peak at intermediate $^{13}$CO fluxes. Higher $^{13}$CO fluxes ($>$500~mJy~km~s$^{-1}$) appear roughly anticorrelated with the HCO$^+$, N$_2$H$^+$, and HCN fluxes, but such a trend is difficult to establish with only 3--4 sources in this range. There is no trend in the lower $^{13}$CO fluxes. The only clear trend we find is a positive correlation between HCO$^+$ fluxes and both N$_2$H$^+$ and HCN fluxes (shown in the last column of Fig.~\ref{fig:obsFlux}). 

In Figure~\ref{fig:obsFluxRatio}, we compare the measured HCO$^+$, N$_2$H$^+$, and HCN fluxes to the derived gas-to-dust mass ratios for our source sample from \cite{2017AA...599A.113M}. These gas-to-dust mass ratios (here referred to as M$\mathrm{_{gas[CO]}}$/M$\mathrm{_{dust}}$) are estimated based on comparison of the sub-mm and $^{13}$CO fluxes to a large model grid assuming interstellar volatile abundances. In cases where C$^{18}$O emission is also detected, the disk gas masses are constrained by both CO isotopologue fluxes. In general, the fluxes of all three molecular species, HCO$^+$, N$_2$H$^+$, and HCN, decrease as M$\mathrm{_{gas[CO]}}$/M$\mathrm{_{dust}}$ increases. But the ranges in the gas-to-dust ratio estimates for each source are large and the trend is not strictly linear. The lowest M$\mathrm{_{gas[CO]}}$/M$\mathrm{_{dust}}$ ratio corresponds to the highest flux for HCO$^+$, N$_2$H$^+$, and HCN. It is clear that Sz~129 stands out in this sample. In addition, HCO$^+$, N$_2$H$^+$, and HCN values remain low for M$\mathrm{_{gas[CO]}}$/M$\mathrm{_{dust}} \gtrsim$10. At intermediate values of M$\mathrm{_{gas[CO]}}$/M$\mathrm{_{dust}}$, there is a range of intermediate to low HCO$^+$, N$_2$H$^+$, and HCN fluxes. We also compare the observed flux ratios of HCO$^{+}$/$^{13}$CO, HCN/$^{13}$CO, and N$_2$H$^+\!$/$^{13}$CO with M$\mathrm{_{gas[CO]}}$/M$\mathrm{_{dust}}$ in Figure~\ref{fig:obsFluxRatio}. The results are similar, with a generally decreasing but not strictly linear trend.

To further examine the robustness of the correlations (or lack thereof) found between source parameters and our measured spectral line fluxes, we use the linear regression method of \cite{2007ApJ...665.1489K}. This method uses a Bayesian approach to account for uncertainties and upper limits in astronomical data. Using the Python package \texttt{linmix}\footnote{\url{https://github.com/jmeyers314/linmix}}, we compute the linear correlation coefficient for each pair of parameters for the entire set of seven sources. The median correlation coefficient for each set is shown in Figure~\ref{fig:stats} with error bars representing one standard deviation. By definition, correlation coefficients range from -1, indicating the strongest possible linear anticorrelation, to 1, indicating the strongest possible linear correlation. A value around zero indicates that no linear correlation is present between the two variables. The top row of Fig.~\ref{fig:stats} presents the correlations coefficients when taking into account the full sample. Here we see strong correlations between the HCO$^+$ and N$_2$H$^+$ and HCN fluxes. We also see a moderate anticorrelation between M$\mathrm{_{gas[CO]}}$/M$\mathrm{_{dust}}$ and our HCO$^+$, N$_2$H$^+$, and HCN fluxes. All other pairs of parameters show little to no correlation. 

Next we test the robustness of these trends and check for potential outliers by leaving out one source at a time and computing the correlation coefficient for the remaining six sources. In all cases, the correlation coefficients between our measured line fluxes and stellar mass, stellar luminosity, and $^{13}$CO flux are close to zero. For the sub-mm continuum flux and mass accretion rate, excluding any source except for Sz~129 (source~7) results in a slight positive correlation between this quantity and our measured line fluxes. This is particularly true when when Sz~68 (source~3) is excluded. Excluding Sz~129 causes this correlation to disappear entirely, suggesting that this source alone is driving the apparent trends with sub-mm continuum flux and mass accretion rate. The slight anticorrelation between our observed molecular line fluxes and M$\mathrm{_{gas[CO]}}$/M$\mathrm{_{dust}}$ is more robust, remaining regardless of which disk is excluded from the sample. Nevertheless, given the relatively large standard deviation in the correlation coefficients, the possible anticorrelation here is not conclusive. The only trend that we tentatively confirm is that the N$_2$H$^+$ and HCN fluxes are positively correlated with HCO$^+$. The median correlation coefficient is 0.88 with a standard deviation of 0.33 for N$_2$H$^+$ and HCO$^+$ and 0.89 with a standard deviation of 0.31 for HCN and HCO$^+$. Possible implications of these findings are further discussed in Section~\ref{section:discussion1}. 

\subsection{Comparison with Model Predictions}

\begin{figure}
\includegraphics[width=\linewidth]{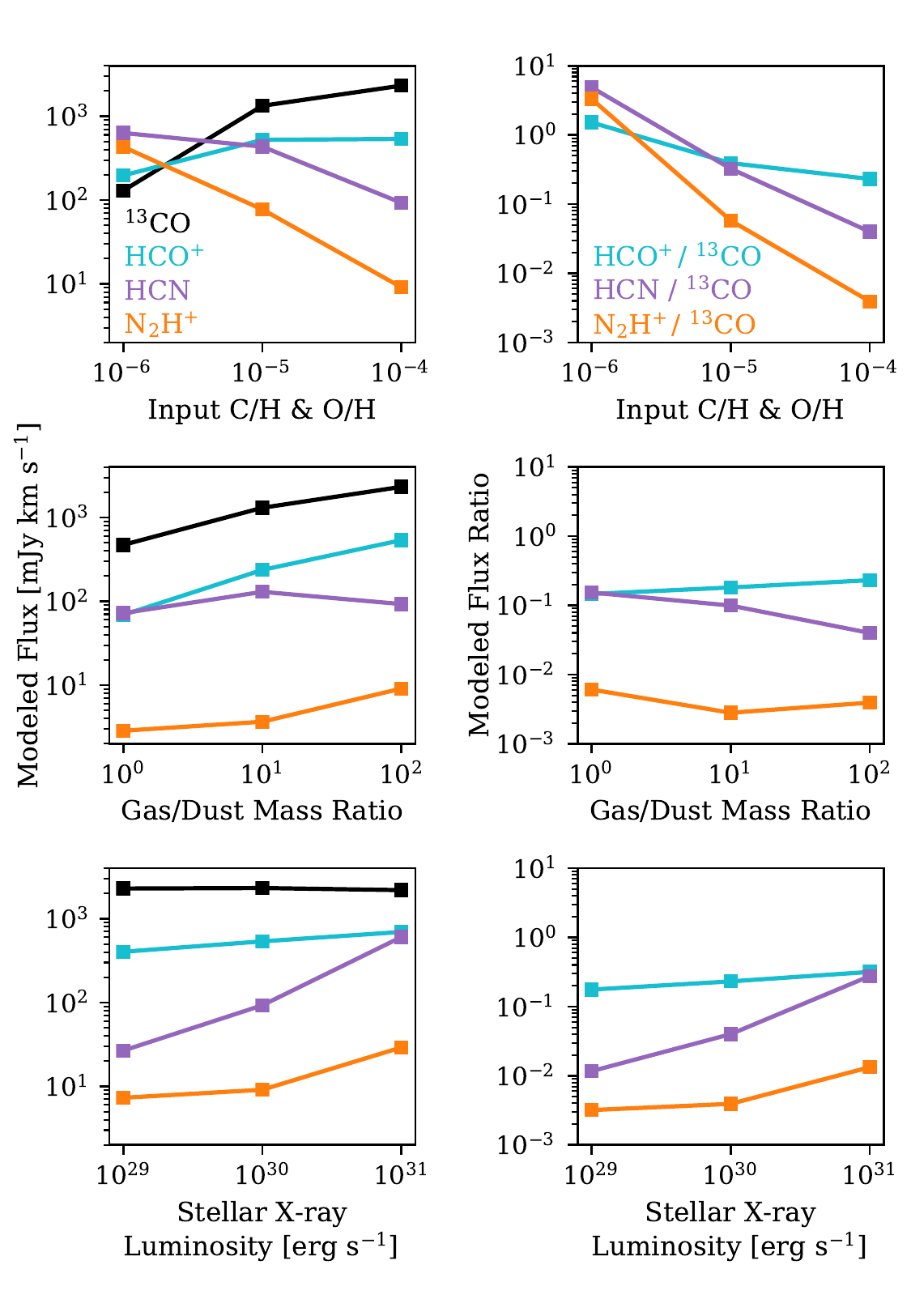}
\caption{Modeled $^{13}$CO, HCO$^+$, HCN, and N$_2$H$^+$ $J$~=~3--2 fluxes (first column) and ratios relative to $^{13}$CO (second column) compared for changing disk parameters. Line fluxes assume a disk inclination of 60$\degr$ and a distance of 160~pc.}
\label{fig:model}
\end{figure}

For comparison with our observations, we simulate $^{13}$CO, HCO$^+$, N$_2$H$^+$, and HCN $J$~=~3--2 fluxes for the series of physical-chemical disk models described in Section~\ref{section:methods3}. The model is not meant to reproduce any one of the observed disks in our sample but instead to demonstrate general expectations for a disk around a T~Tauri star and the dependence of line emission for our selected molecular species on particular disk parameters. Based on our model results shown in Figure~\ref{fig:model}, for increasing gas-to-dust mass ratios we expect the fluxes of all four molecular species to increase or, in the case of HCN, remain roughly constant. In contrast, the trends diverge for increasing volatile abundances of C/H and O/H. These findings imply that the degeneracy between gas mass and volatile abundances can be broken with multi-molecular observations. For example, fluxes of $^{13}$CO and HCO$^+$ increase as C/H and O/H increase, whereas N$_2$H$^+$ and HCN decrease. In the case of N$_2$H$^+$, the decrease in flux is particularly significant as previously explored by \citet{2019ApJ...881..127A}. The $^{13}$CO emission modeled with LIME is optically thick in all cases, except when C/H and O/H are depleted to 10$^{-6}$. Fluxes for the other molecular species become optically thick at values above a few hundred mJy~km~s$^{-1}$.

While our main focus is disentangling gas mass and volatile abundance effects, we further test the effects of increasing disk ionization by varying the stellar X-ray luminosity. A full exploration of sources and possible distributions of ionization throughout the disk is beyond the scope of this work. Such variations will undoubtedly result in different trends for the fluxes of our selected molecular species, but here we demonstrate one example case. In our models, HCO$^+$, N$_2$H$^+$, and HCN fluxes increase with increasing stellar X-ray luminosity by factors of 2, 4, and 20, respectively. Meanwhile, $^{13}$CO fluxes slightly decrease by 4\%. 

As shown in the second column of Fig.~\ref{fig:model}, these differing behaviors between molecular species result in different trends in the flux ratios relative to $^{13}$CO for the cases of volatile (C/H and O/H) depletion and gas depletion. In particular, the N$_2$H$^+\!$/$^{13}$CO flux ratio is expected to decrease significantly as volatile abundances increase, whereas it remains largely constant as gas-to-dust mass ratios increase. In the case of increasing stellar X-ray luminosities, N$_2$H$^+\!$/$^{13}$CO and HCN/$^{13}$CO flux ratios increase while HCO$^+\!$/$^{13}$CO remains relatively constant.

To first order, the observed trends deviate from the model predictions, suggesting there are other important parameters controlling the chemistry that vary across the sample. For example, we find a range of over an order of magnitude in N$_2$H$^+\!$/$^{13}$CO and HCN/$^{13}$CO flux ratios, and a factor of $\sim$5$\times$ in HCO$^{+}\!$/$^{13}$CO in our observations of the Lupus disks. This makes our observations inconsistent with simply changing gas-to-dust mass ratios (middle row of Fig.~\ref{fig:model}), which produces relatively constant flux ratios. But unlike the case of decreasing C/H and O/H (top row of Fig.~\ref{fig:model}), our observed HCO$^{+}$ fluxes do not follow the behavior of $^{13}$CO as would be expected if their chemistries were linked. Similar to the case of differing stellar X-ray luminosities (bottom row of Fig.~\ref{fig:model}), we observe positively correlated trends in the behavior of HCO$^{+}$, HCN, and N$_2$H$^+$. However, even when the stellar X-ray luminosity is increased substantially to a level of 10$^{31}$~erg~s$^{-1}$, the modeled $^{13}$CO fluxes do not vary significantly or decrease enough to approach the range of the observed values.  
 
 \subsection{Disk Gas Masses from Multi-molecular Observations}

\begin{figure}
\includegraphics[width=\linewidth]{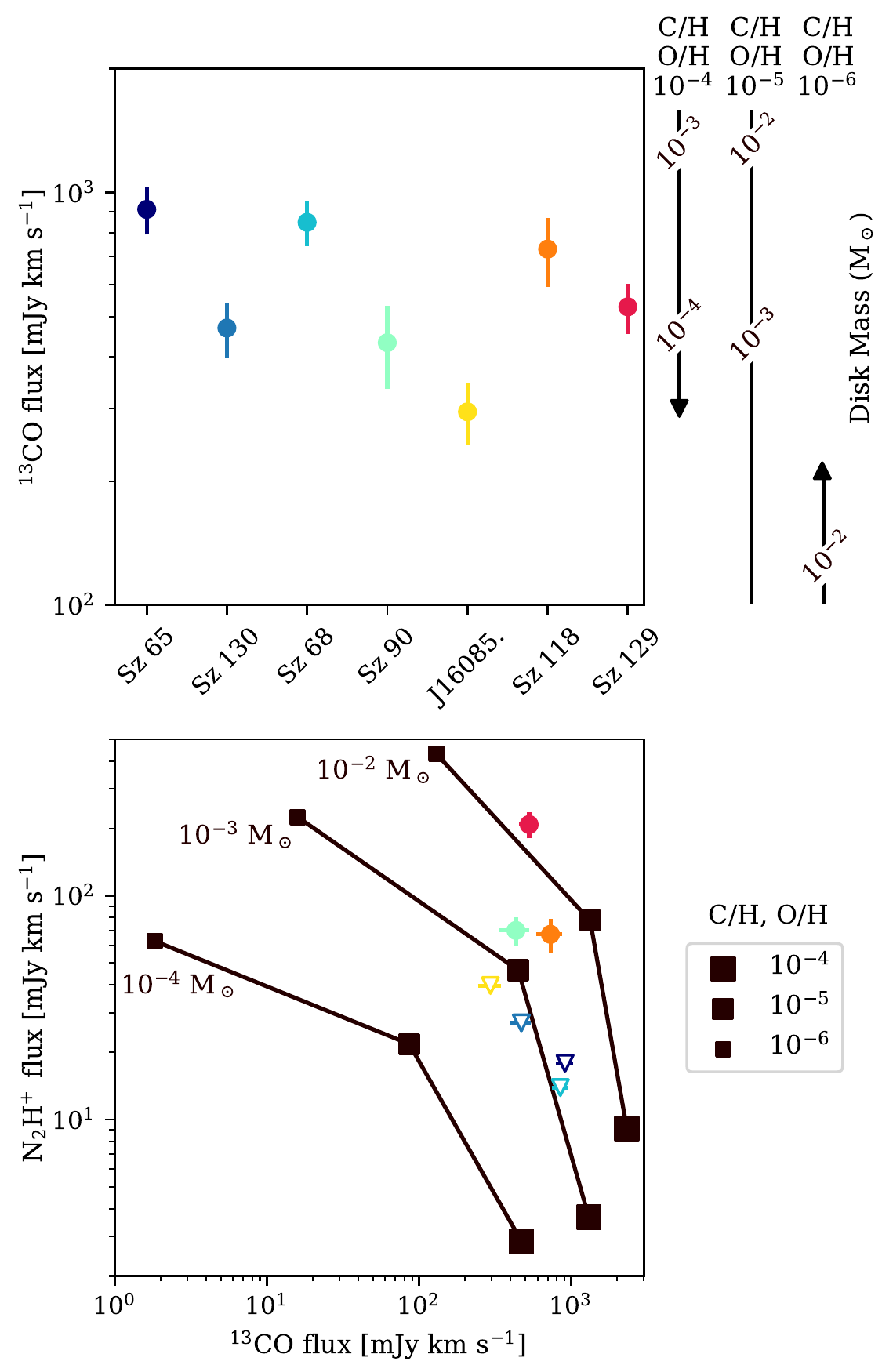}
\caption{Comparison of disk gas mass estimates for the sample of Lupus disks using $^{13}$CO fluxes alone (top panel) vs.~a combination of $^{13}$CO and N$_2$H$^+$ fluxes (bottom panel). To the right of the top panel, the corresponding disk gas masses for a given $^{13}$CO flux are indicated for three different assumed values of C/H and O/H. The uncertainty in disk gas mass is large given the lack of constraints on the elemental abundances in the gas. In the bottom panel, the squares indicate the N$_2$H$^+$ and $^{13}$CO fluxes for different combinations of disk gas masses and C/H, O/H abundances in our models. The various colored points indicate the observed line fluxes, which are scaled to the distance of 160~pc assumed in the models. Downward triangles represent upper limits. The same colors as used in the plots above are used to distinguish the seven different sources.}
\label{fig:mass}
\end{figure}

While our model grid does not cleanly predict any one trend in the data with a single dependent variable, we can examine the data in the context of the full set of models to place constraints on the disk masses in the sample. The traditional approach of using CO isotopologue fluxes to estimate disk gas masses is demonstrated in the top panel of Figure~\ref{fig:mass}, where we highlight the degeneracy with volatile abundance that has been previously reported \citep{2016ApJ...828...46A,2017AA...599A.113M,2017ApJ...844...99L}. As shown, CO isotopologue fluxes alone can be associated with a wide range of orders of magnitude in mass depending on the C/H and O/H abundances assumed. If we incorporate one additional constraint, namely the N$_2$H$^+$ flux, the translation to mass fills a plane as indicated in the bottom panel of Figure~\ref{fig:mass}.

By combining N$_2$H$^+$ and $^{13}$CO measurements we begin to break the current degeneracy involved in disk gas mass estimations. Using the set of models described above, we find that relatively large disk gas masses combined with carbon and oxygen depletion are needed to reproduce the N$_2$H$^+$ fluxes detected in three of the Lupus disks (bottom panel of Fig.~\ref{fig:mass}). This is in agreement with previous findings that suggest that at least some subset of the Lupus disks have CO-depleted gas \citep{2016AA...591L...3M,2019AA...631A..69M}. The new mass estimates of $>$10$^{-3}$--10$^{-2}$~M$_\odot$ are 10--100$\times$ higher than previous ranges for these sources based on CO isotopologues when interstellar C/H and O/H abundances are used \citep{2017AA...599A.113M}. Other unconstrained parameters, including disk ionization rates, still cause uncertainty in these values. Yet we find that the scenario is more complex than simply varying gas masses alone to explain the range of M$\mathrm{_{gas[CO]}}$/M$\mathrm{_{dust}}$ ratios seen in the Lupus disk population because of the strength of the N$_2$H$^+$ emission observed in some disks with low M$\mathrm{_{gas[CO]}}$/M$\mathrm{_{dust}}$.  

\section{Discussion}\label{section:discussion}
We now explore the implications of the results obtained through our observational and modeling analyses. First we discuss the trends in the observed molecular fluxes. Then we evaluate our ability to constrain H$_2$ gas masses of disks. 

\subsection{Trends in HCO$^+$, N$_2$H$^+$, and HCN Fluxes}\label{section:discussion1}

The first trend noted in our observations is that the relationship between the observed HCO$^{+}$ and CO isotopologue fluxes is different from that predicted in our models. Protonation of CO by H$_3^+$ is one of the main formation pathways of HCO$^{+}$ \citep[e.g.,][]{2012ApJ...747..114W,2015ApJ...799..204C,2018A&A...616A..19A}. As shown in Figure~\ref{fig:model}, the HCO$^{+}$ flux is expected to increase alongside the $^{13}$CO flux for both increasing C/H, O/H abundances (top row) and gas-to-dust mass ratios (middle row). We expected HCO$^{+}$ to indirectly track the CO content of the disk. In contrast, as Fig.~\ref{fig:obsFlux} shows, the highest $^{13}$CO (and by association, C$^{18}$O) fluxes correspond to some of the lowest HCO$^{+}$ fluxes. 

This lack of correlation between observed HCO$^{+}$ and $^{13}$CO fluxes could be because their emission is optically thick and tracking the (unconstrained) temperature structure of the disk. Based on optical depth predictions from LIME, we expect $^{13}$CO emission may be optically thick in all seven disks, while HCO$^+$ is predicted to be optically thick for fluxes above $\sim$250~mJy~km~s$^{-1}$ (about half of our sample). The observed ratios of C$^{18}$O/$^{13}$CO fluxes further support the interpretation that the $^{13}$CO emission is optically thick. The optically thick line fluxes may be tracking temperature conditions in the disk, in which case, the observed difference in HCO$^{+}$ and $^{13}$CO fluxes could suggest that they are predominately originating from different regions of the disk. Alternatively, the lack of correspondence between the observed HCO$^{+}$ and $^{13}$CO fluxes could be related to sources of ionization and how they are distributed throughout the disk.

The second trend apparent in the observed fluxes is the tentative correlation between HCO$^{+}$, HCN, and N$_{2}$H$^{+}$. It is not clear why this is the case. It could be that these three molecular species are more accurately tracking the H$_2$ mass of the disk whereas the sub-mm and CO isotopologue fluxes are not. In that case, we would expect the correlation to extend to other molecular species that are sensitive to the bulk disk mass as well. 

Another possibility is that HCO$^{+}$, HCN, and N$_{2}$H$^{+}$ are all influenced by the same sources of ionization in the disk. This stands to reason because HCO$^{+}$ and N$_{2}$H$^{+}$ are both molecular ions. Furthermore, their dominant formation pathways throughout much of the H$_2$-rich disk gas involve H$_3^+$. In the case of HCN, high X-ray ionization rates leading to the breaking up of CO via He$^+$ combined with H$_2$O freezeout removing oxygen from the gas in the outer disk could produce high C/O in the gas. HCN abundances are enhanced at high C/O. Such enhancement is also expected for C$_2$H, which was detected in Sz~129 and an upper limit found for Sz~68 \citep{2019AA...631A..69M}. While a full exploration of the effects of varying C/O ratios is beyond the scope of this work, we test different values of C/O in a 0.01~M$_{\odot}$ disk by adding the necessary amount of C$^{+}$ to the initial abundances with C/H and O/H on the order of 10$^{-6}$. HCN fluxes are the most affected, increasing 1.8$\times$ for C/O~=~1 and 2.0$\times$ for C/O~=~1.5 relative to the initial fluxes for a C/O of 0.6. Over this range, $^{13}$CO fluxes increase by 1.5--1.6$\times$, HCO$^{+}$ increase by 1.3--1.4$\times$, and N$_2$H$^{+}$ decrease by 1.2--1.4$\times$. 

A third possibility is that these molecular fluxes are sensitive to the disk size. We have not estimated gas radii from the molecules observed here, but \citet{2021AA...649A..19S} provides disk radii from 0.89-mm dust and $^{12}$CO $J$~=~2--1 emission for five of our sources. For these sources, we do not find any trends between our observed HCO$^{+}$, HCN, and N$_{2}$H$^{+}$ fluxes and the CO, dust, or ratio of CO to dust disk radii from \citet{2021AA...649A..19S}.  

\subsection{Disk H$_2$ Mass Estimation}

Identifying accurate gas mass tracers is important to the study of circumstellar disks. \cite{2021arXiv211105833R} recently concluded that C$^{18}$O is an appropriate gas mass tracer for Lupus disks where C$^{18}$O $J$~=~3--2 emission was detected by \cite{2016ApJ...828...46A}, without needing to invoke CO depletion. The upper limits on N$_2$H$^+$ emission for Sz~65 and Sz~68, the two disks in our sample where C$^{18}$O was detected by \cite{2016ApJ...828...46A}, are consistent with a lack of C/H and O/H depletion in these sources. Meanwhile, the N$_2$H$^+$ fluxes seen in Sz~129, Sz~90, and Sz~118 suggest that there are a subset of disks in Lupus for which $^{13}$CO or C$^{18}$O alone would be insufficient gas tracers. 

Our analysis suggests that the H$_2$ masses of Sz~129, Sz~90, and Sz~118 are $>$10--100$\times$ higher than previous estimates \citep{2016ApJ...828...46A,2017AA...599A.113M}. While these values are required to explain the observed N$_2$H$^+$ fluxes, they are not yet well-constrained. Unlike in the work of \citet{2017AA...599A.113M}, here we employ only a small model grid. We make no effort to match the entire set of observed quantities for any of the individual disks and do not explore a sufficient range of parameter space to encapsulate all of the uncertainties. In addition, we selected favorable modeling conditions for N$_2$H$^+$ production: all initial elemental N/H in the form of N$_2$ (with very minor initial amounts of NH$_3$ and HCN), high stellar X-ray luminosities, and increasing amounts of C/H and O/H depletion relative to the canonical interstellar values. Our assumed cosmic-ray ionization rates are sub-interstellar reflecting attenuation within the disk environment \citep{2015ApJ...799..204C, 2021ApJ...912..136S}. Under conditions less favorable for N$_2$H$^+$ production, disk gas masses would need to be even higher to reproduce the observed N$_2$H$^+$ emission. 

In combination, all the observed molecular fluxes could together place further constraints on the elemental composition and ionization of the disk gas. But given additional uncertainties, including the temperature structure of the disk, such an analysis is beyond the scope of this work. Furthermore, the sample of Lupus disks represent a range of stellar luminosities \citep{2019AA...629A.108A}, dust masses \citep{2016ApJ...828...46A, 2017AA...599A.113M}, and disk sizes in continuum and CO emission \citep{2021AA...649A..19S}. The added effects of varying these parameters are also beyond the scope of our current model grid. Future work pursuing various combinations of physical and chemical parameters will require a significantly larger grid of models. 

\cite{2022arXiv220109900T} investigated the use of N$_2$H$^+$ and C$^{18}$O fluxes to constrain the disk gas masses of TW~Hya, GM~Aur, and DM~Tau in comparison to prior values provided by HD. Their disk-specific models calculate N$_2$H$^+$ abundances from a limited chemical network and include cosmic rays as the only ionization source. They also conclude that constraining ionization levels in disks is required to reduce the uncertainty in disk gas mass estimates derived from N$_2$H$^+$. Given the degeneracy between CO depletion and enhanced ionization rates in increasing N$_2$H$^+$ abundances, it is clear that N$_2$H$^+$ is not a suitable \textit{replacement} for CO isotopologues as a gas mass tracer. Instead we suggest that some combination of multiple observable chemical species is necessary to better constrain H$_2$ gas masses of protoplanetary disks.

\section{Conclusions}\label{section:conc}
We present a line survey of seven disks from the Lupus star forming region. Combining our observations of HCO$^+$, N$_2$H$^+$, and HCN with literature CO isotopologue values, we attempt to more holistically constrain the gas masses of these sources in a systematic way. In addition, we search for trends with different known system parameters, such as stellar mass and luminosities, mass accretion rates, and dust masses, and compare to model expectations. We do not see simple trends with any one parameter, suggestive that a combination of physical parameters, some of which are explored in our model grid and some beyond the grid's scope, likely drive the relationships seen in the data. 

The primary observed relationship found is between HCO$^+$, N$_2$H$^+$, and HCN, which are positively correlated with each other, while surprisingly no correlation was found between $^{13}$CO and the other species. To interpret these observations in the context of gas masses, we compare the $^{13}$CO and N$_2$H$^+$ fluxes of our sample to our model grid. We find that combining the lines is useful in measuring disk gas masses and arrive at masses that are substantially higher than previous estimates. Given the dependencies of N$_2$H$^+$ and the assumptions made in the model, the disk masses could be even higher. The full dimensional nature of this grid should be explored in future work. In addition, we also find the following: 

\begin{itemize}
\item Using our physical-chemical models, we are unable to reproduce the measured N$_2$H$^+$ fluxes for Sz~129, Sz~118, or Sz~90 with gas masses below $
\sim$10$^{-3}$--10$^{-2}$ M$_\odot$. In addition, significant CO depletion is required, even at higher disk masses. This analysis provides a unique line of evidence, in addition to accretion rates \citep{2016AA...591L...3M} and hydrocarbon emission \citep{2019AA...631A..69M}, that the CO-based gas masses are underestimating the total H$_2$ mass for at least some of the disks in the Lupus star-forming region. Future observations of a larger number of disks using multiple molecular gas tracers are required to provide meaningful statistics across this and other disk populations. 

\item HCO$^{+}$ fluxes do not correlate with the $^{13}$CO (or the limited number of detected C$^{18}$O) fluxes. If HCO$^{+}$ is not directly tracking the CO content of the disk, then there must be another factor, such as H$_2$ mass, ionization, or temperature, that drives the trends in its emission.

\item Sz~129, the source with the lowest M$\mathrm{_{gas[CO]}}$/M$\mathrm{_{dust}}$ in our sample, stands out from the rest. This source has the highest 0.890-mm, HCO$^+$, N$_2$H$^+$, and HCN fluxes in addition to the highest flux ratios relative to $^{13}$CO. These line fluxes and flux ratios are about 1.5--3$\times$ higher than the next highest value. Meanwhile, the disks with the highest $^{13}$CO and C$^{18}$O fluxes and CO-to-dust ratios are among those with the lowest HCO$^+$, N$_2$H$^+$, and HCN fluxes. The opposing behavior of CO isotopologue and N$_2$H$^+$ fluxes at the extremes is consistent with the range of M$\mathrm{_{gas[CO]}}$/M$\mathrm{_{dust}}$ being at least partially caused by CO depletion. But the lack of such trends at intermediate M$\mathrm{_{gas[CO]}}$/M$\mathrm{_{dust}}$ values suggests that other factors are also influencing the line fluxes and flux ratios.
\end{itemize}

Accurate measurements of disk masses are crucial for developing our understanding of when and where planet formation can occur. The investigation of molecular species beyond CO, including non-carbon-bearing volatiles like N$_2$H$^+$, will be instrumental in the development of future experiments aimed at identifying gas-rich disks in large surveys. This is particularly pertinent if CO-depletion is a widespread phenomenon among disk populations even at young ages, as the faint CO emission found in current surveys may suggest \citep[e.g.,][]{2016ApJ...828...46A,2017ApJ...844...99L}.

\software {Astropy \citep{astropy:2013, astropy:2018}, CASA \citep{2007ASPC..376..127M}, linmix (\url{https://github.com/jmeyers314/linmix}), Matplotlib \citep{matplotlib}, Numpy \citep{numpy}, parallel \citep{tange_ole_2018_1146014}}

\acknowledgments We thank the anonymous referee for their helpful feedback and suggestions. D.E.A. acknowledges support from the Virginia Initiative on Cosmic Origins (VICO) Postdoctoral Fellowship. L.I.C. gratefully acknowledges support from the David and Lucille Packard Foundation, the Virginia Space Grant Consortium, Johnson \& Johnson's WiSTEM2D Award, and NSF AAG grant number AST-1910106. This paper makes use of the following ALMA data: ADS/JAO.ALMA\#2019.1.01135.S. ALMA is a partnership of ESO (representing its member states), NSF (USA) and NINS (Japan), together with NRC (Canada), MOST and ASIAA (Taiwan), and KASI (Republic of Korea), in cooperation with the Republic of Chile. The Joint ALMA Observatory is operated by ESO, AUI/NRAO and NAOJ. We also acknowledge the University of Virginia's Rivanna computing cluster, which was used to run some of the models used in this work. 

\bibliography{bibliography2021}

\begin{figure*}
  \includegraphics[width=\linewidth]{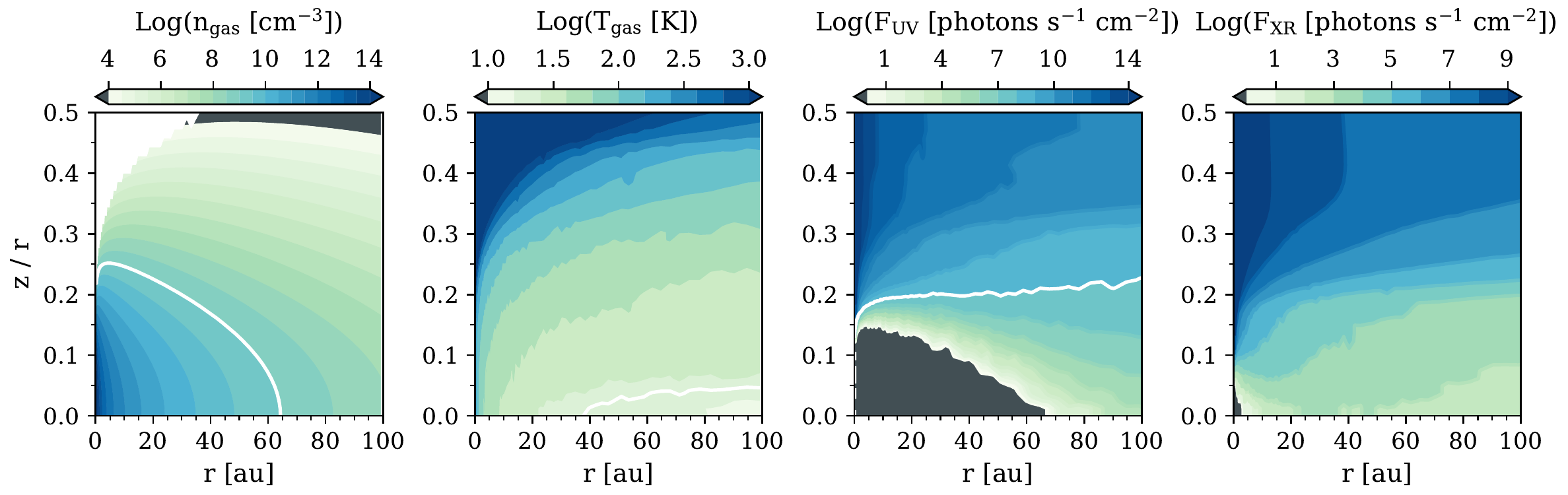}
  \caption{The physical structure of the modeled fiducial 10$^{-2}$~M$_\odot$ disk including the gas density, gas temperature, and UV continuum and X-ray radiation fields. Contours indicate the values of 10$^{9}$~cm$^{-3}$, 20~K, and G$_0$ \citep{1968BAN....19..421H} for reference.}
  \label{fig:env}
\end{figure*}

\begin{figure}
\begin{minipage}{.49\textwidth}
  \includegraphics[width=\linewidth]{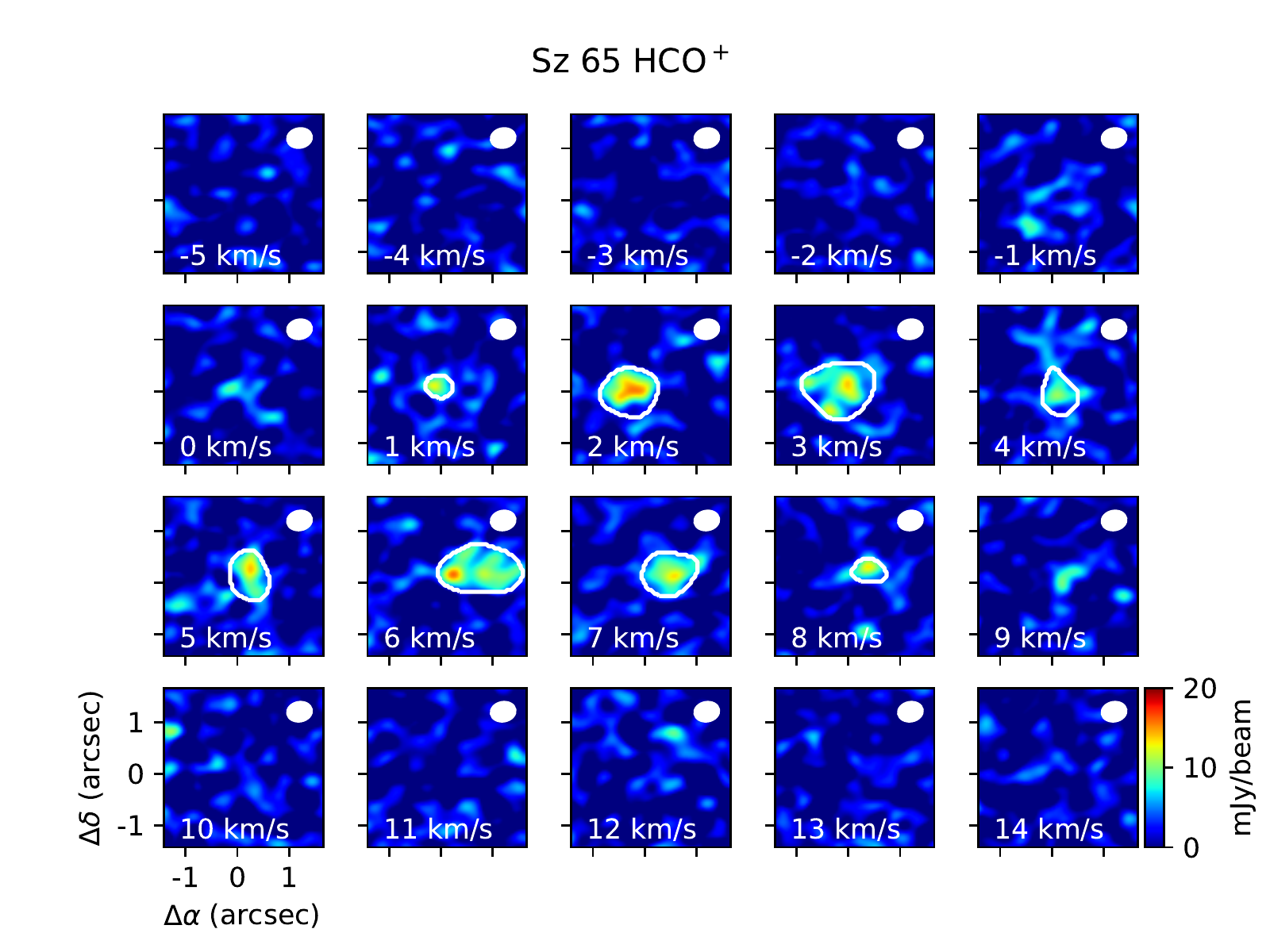}
  \includegraphics[width=\linewidth]{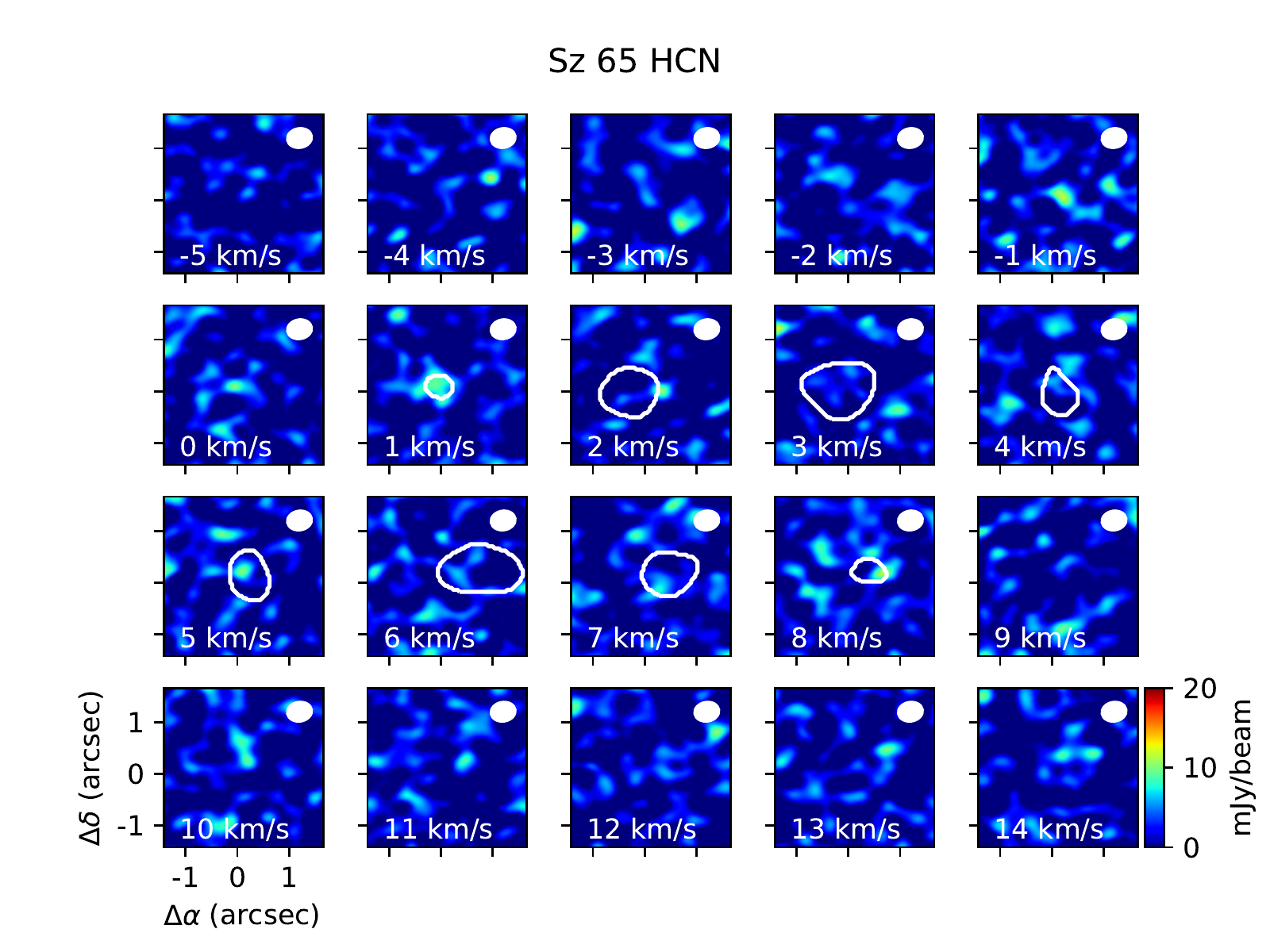}
  \includegraphics[width=\linewidth]{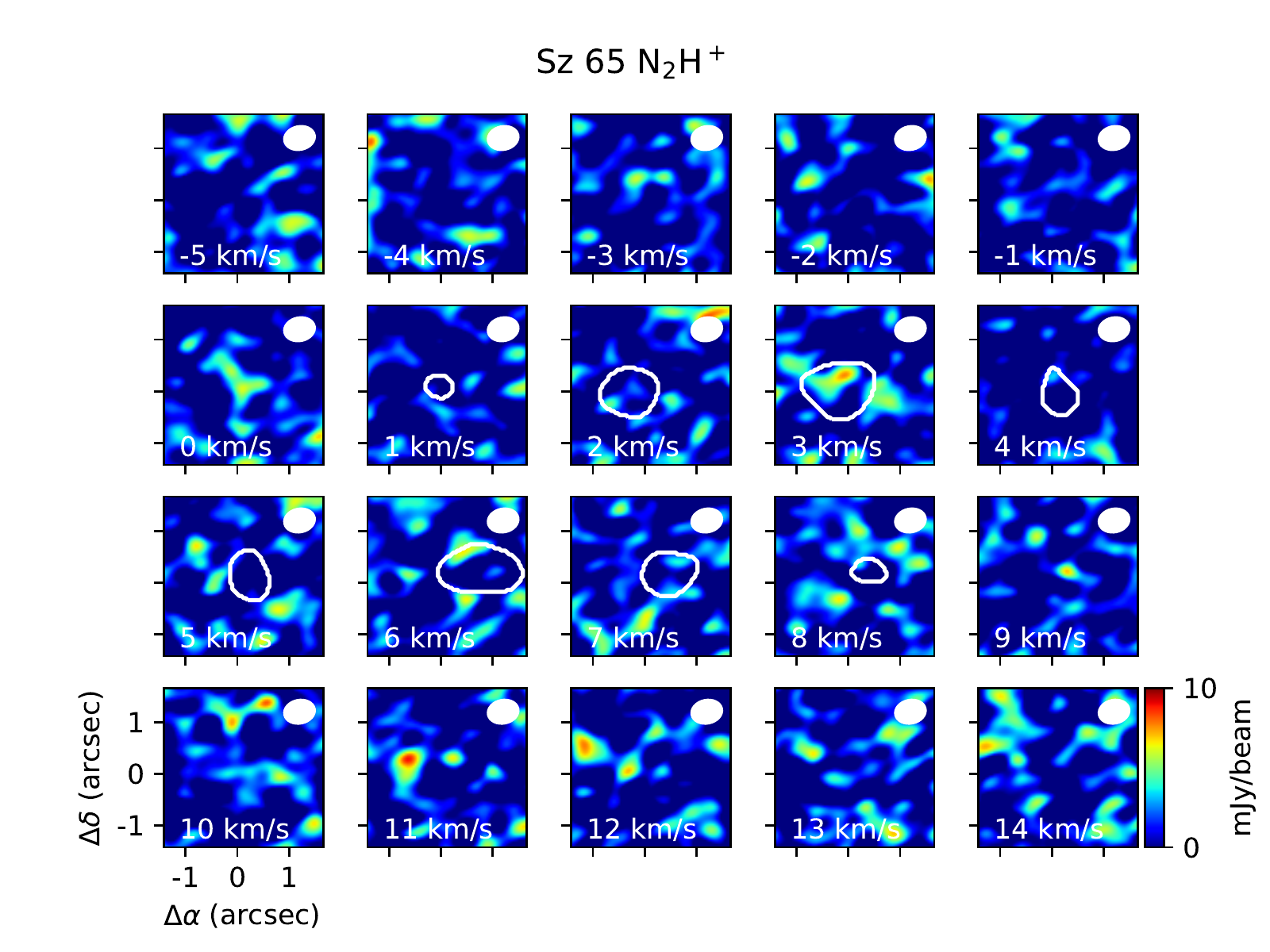}
  \captionsetup{width=0.95\linewidth}
  \caption{Channel maps of HCO$^+$, HCN, and N$_2$H$^+$ line emission for Sz~65. Contours outline the mask generated based on the HCO$^+$ emission. Tick marks indicate 1\arcsec. The beam is shown in the upper right. }
  \label{fig:maps1}
\end{minipage}%
\begin{minipage}{.49\textwidth}
  \includegraphics[width=\linewidth]{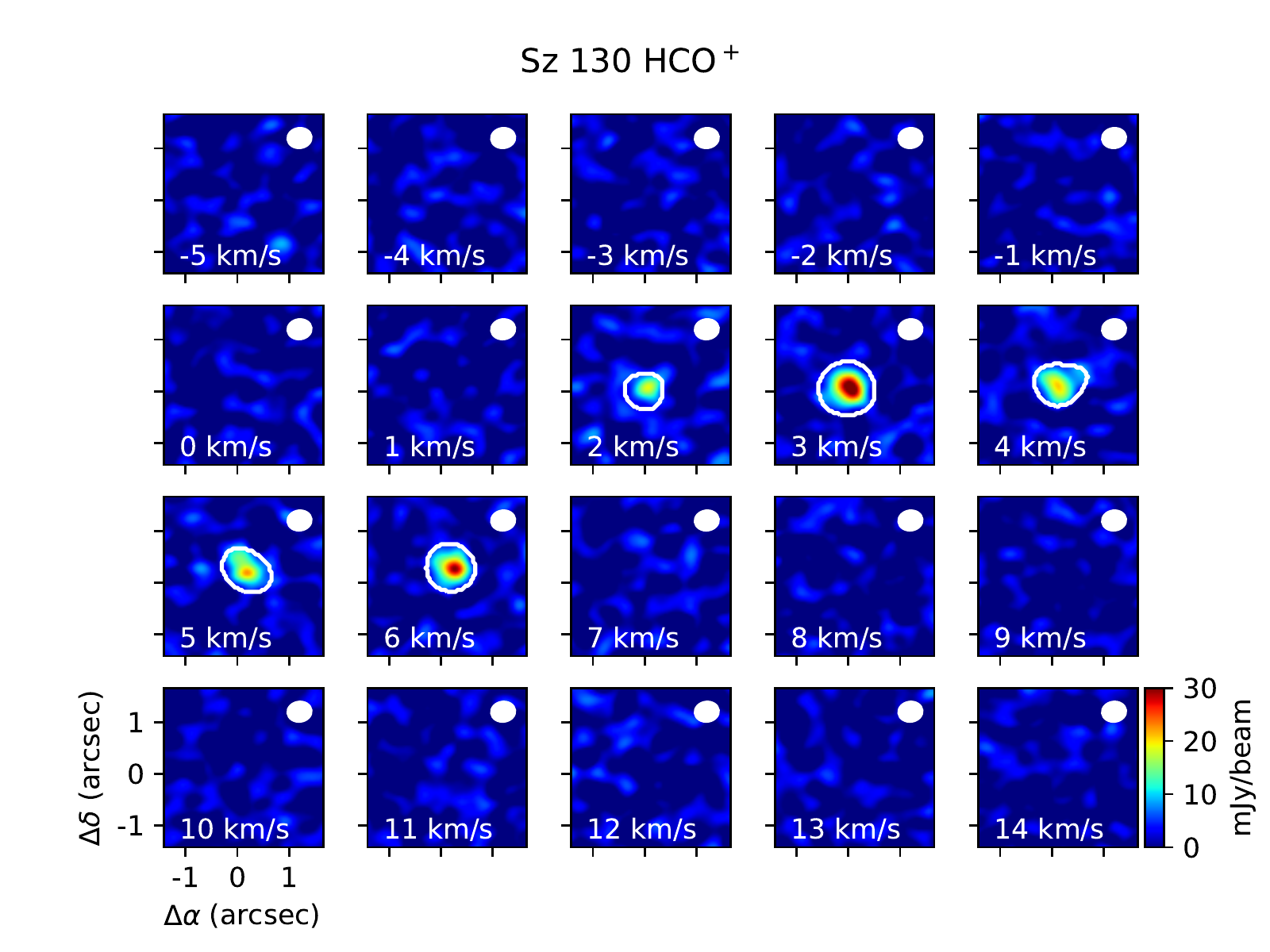}
  \includegraphics[width=\linewidth]{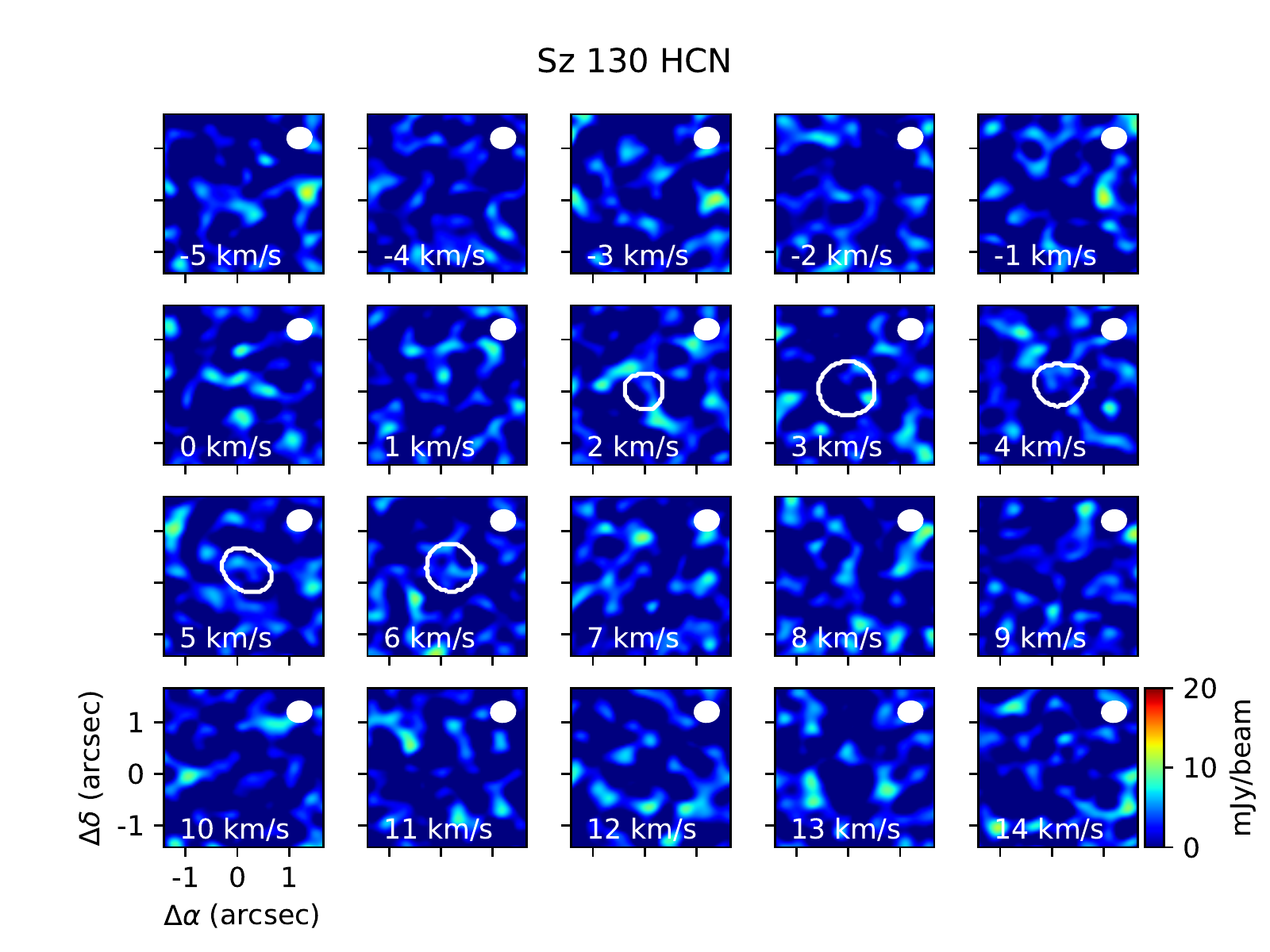}
  \includegraphics[width=\linewidth]{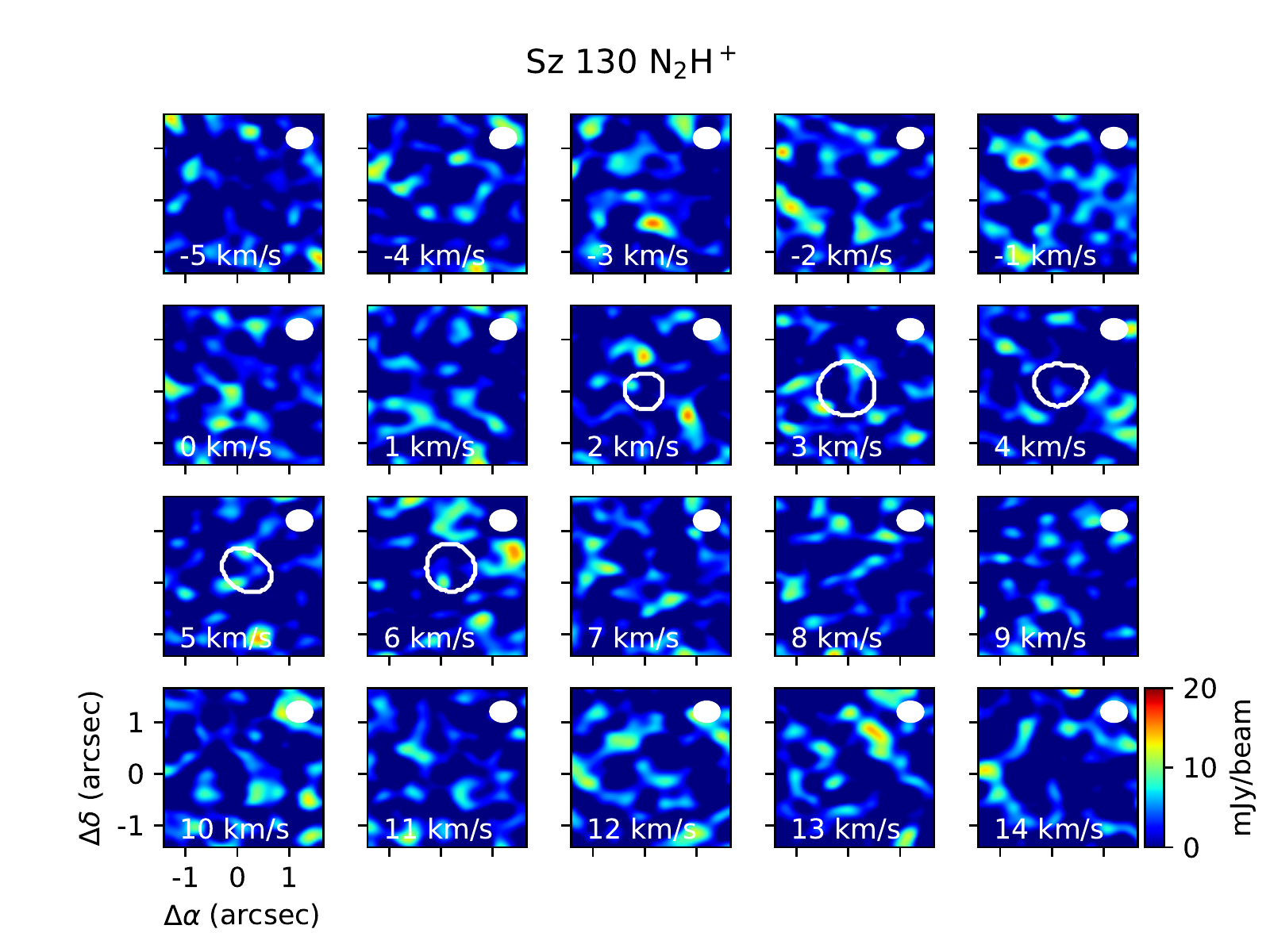}
  \captionsetup{width=0.95\linewidth}
  \caption{Channel maps of HCO$^+$, HCN, and N$_2$H$^+$ line emission for Sz~130. Contours outline the mask generated based on the HCO$^+$ emission. Tick marks indicate 1\arcsec. The beam is shown in the upper right.
}
\label{fig:maps2}
\end{minipage}
\end{figure}

\begin{figure}
\begin{minipage}{.49\textwidth}
  \includegraphics[width=\linewidth]{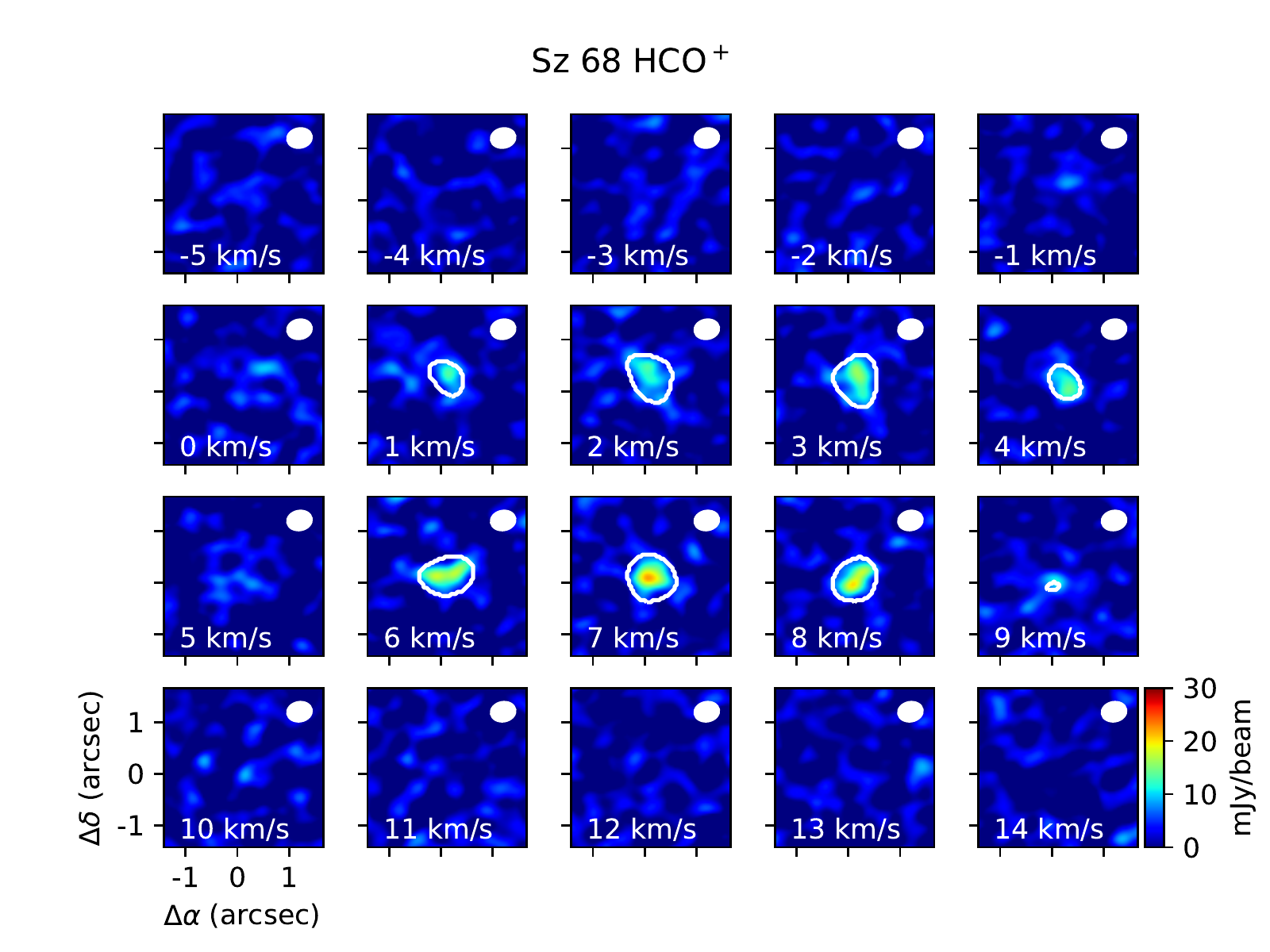}
  \includegraphics[width=\linewidth]{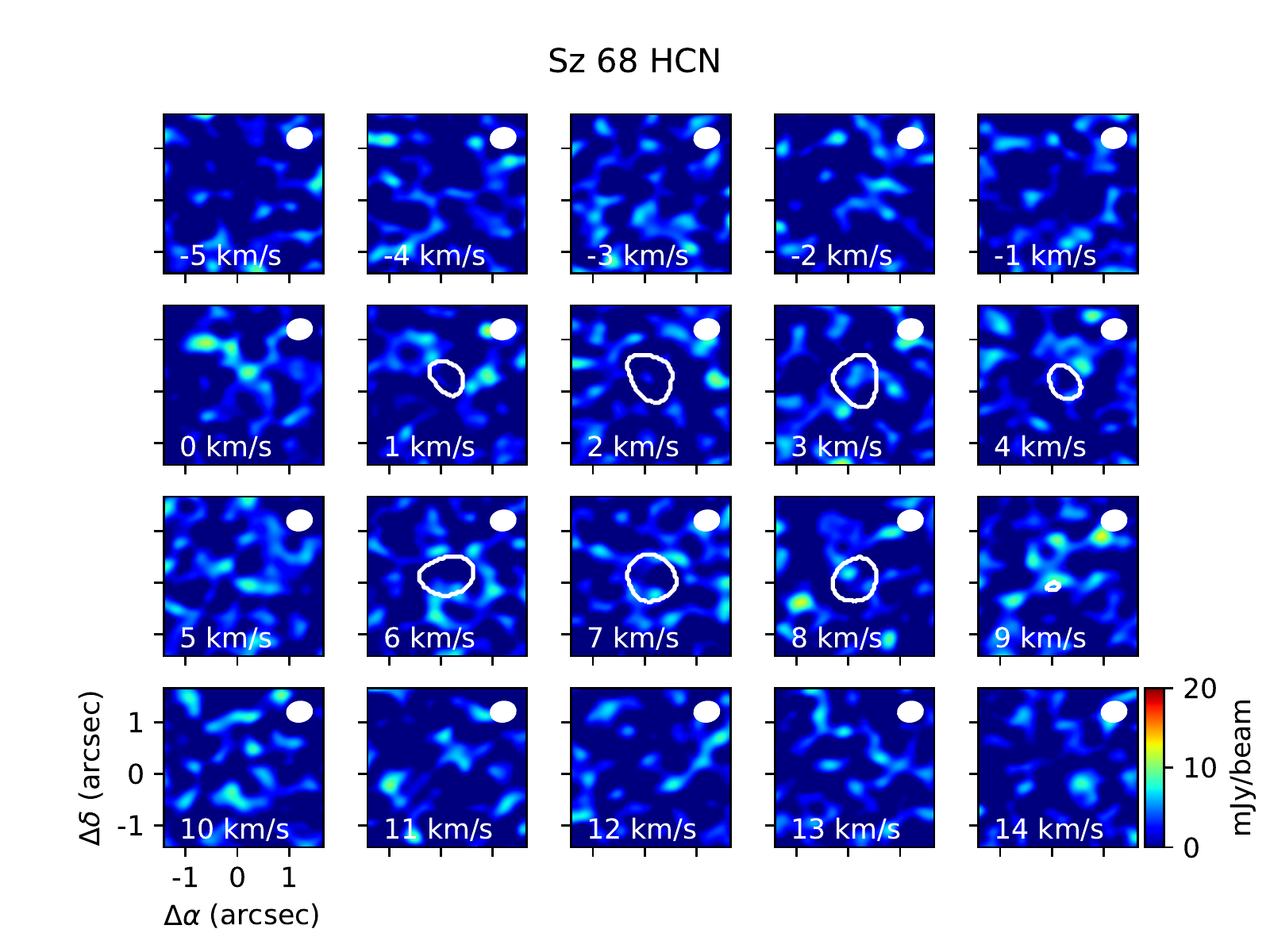}
  \includegraphics[width=\linewidth]{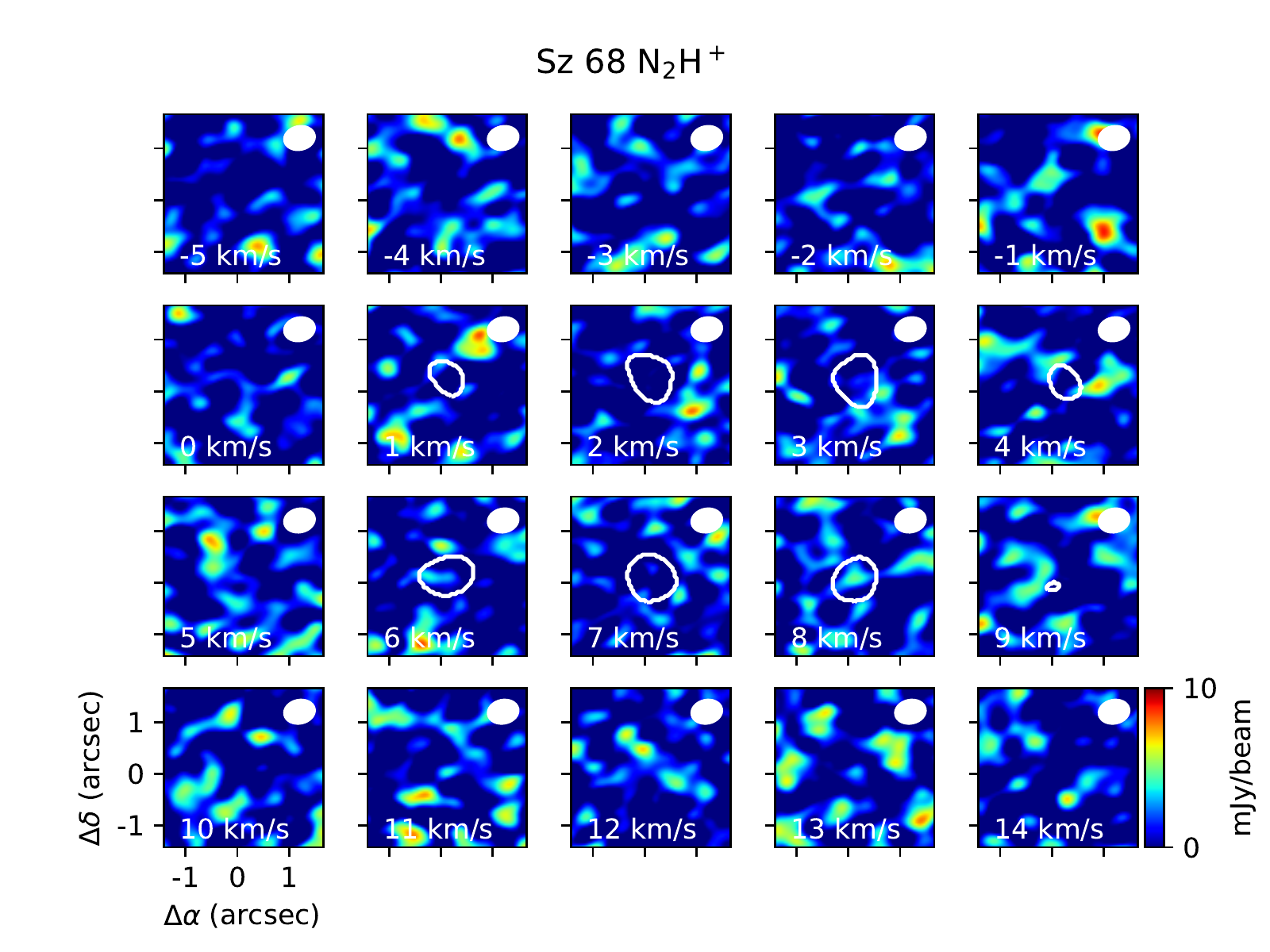}
  \captionsetup{width=0.95\linewidth}
  \caption{Channel maps of HCO$^+$, HCN, and N$_2$H$^+$ line emission for Sz~68. Contours outline the mask generated based on the HCO$^+$ emission. Tick marks indicate 1\arcsec. The beam is shown in the upper right. }
  \label{fig:maps3}
\end{minipage}%
\begin{minipage}{.49\textwidth}
  \includegraphics[width=\linewidth]{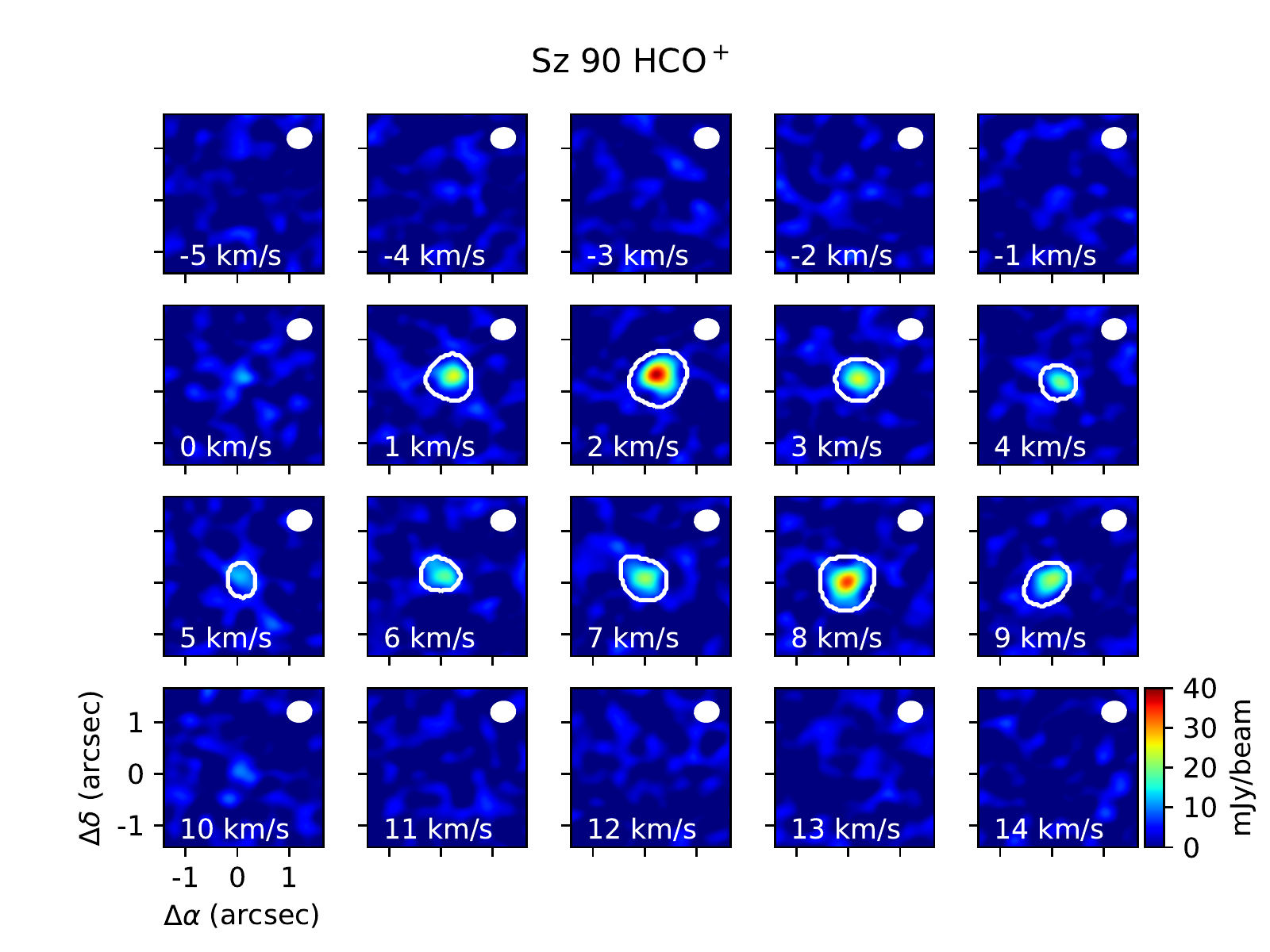}
  \includegraphics[width=\linewidth]{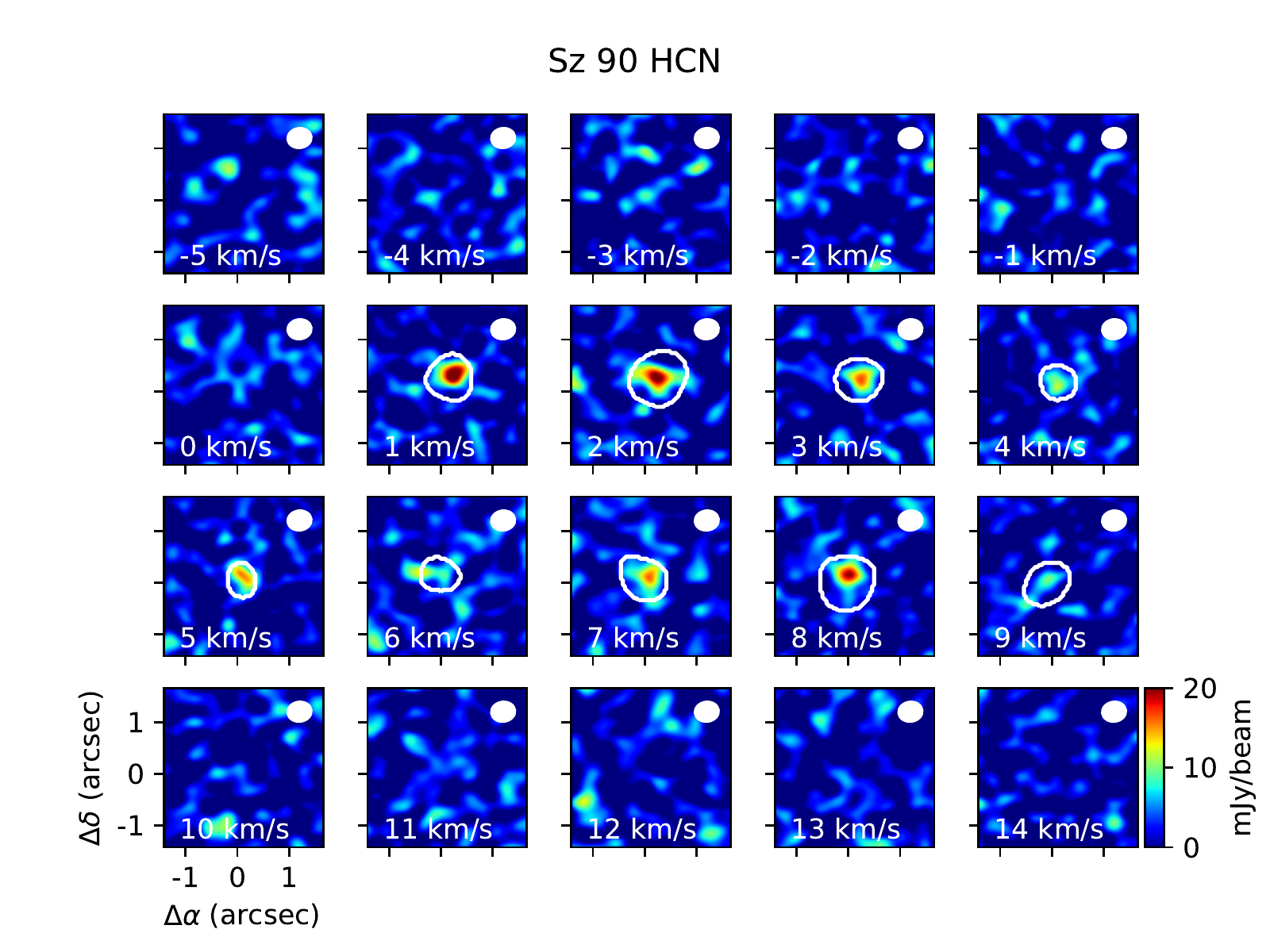}
  \includegraphics[width=\linewidth]{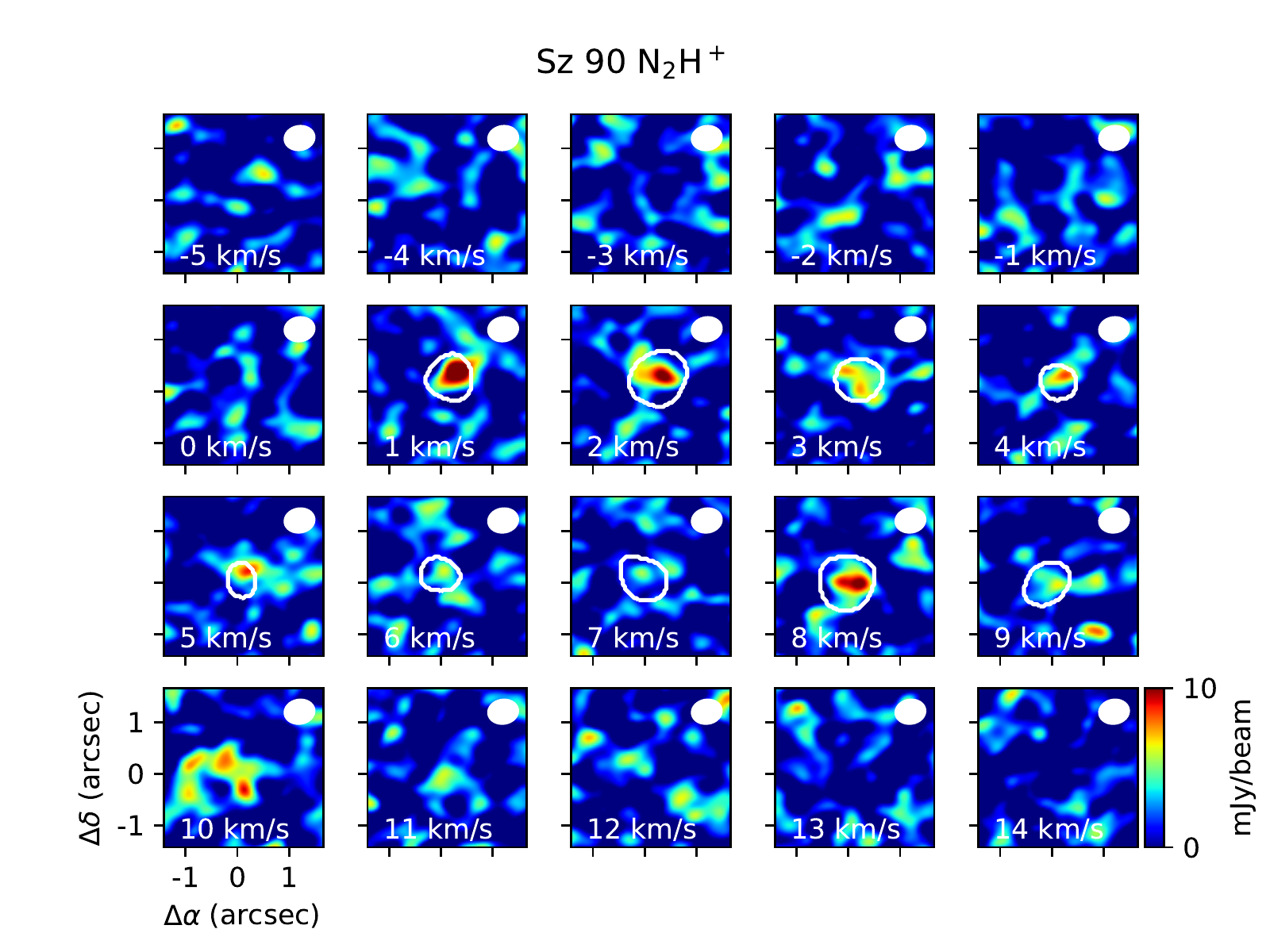}
  \captionsetup{width=0.95\linewidth}
  \caption{Channel maps of HCO$^+$, HCN, and N$_2$H$^+$ line emission for Sz~90. Contours outline the mask generated based on the HCO$^+$ emission. Tick marks indicate 1\arcsec. The beam is shown in the upper right.
}
\label{fig:maps4}
\end{minipage}
\end{figure}

\begin{figure}
\begin{minipage}{.49\textwidth}
  \includegraphics[width=\linewidth]{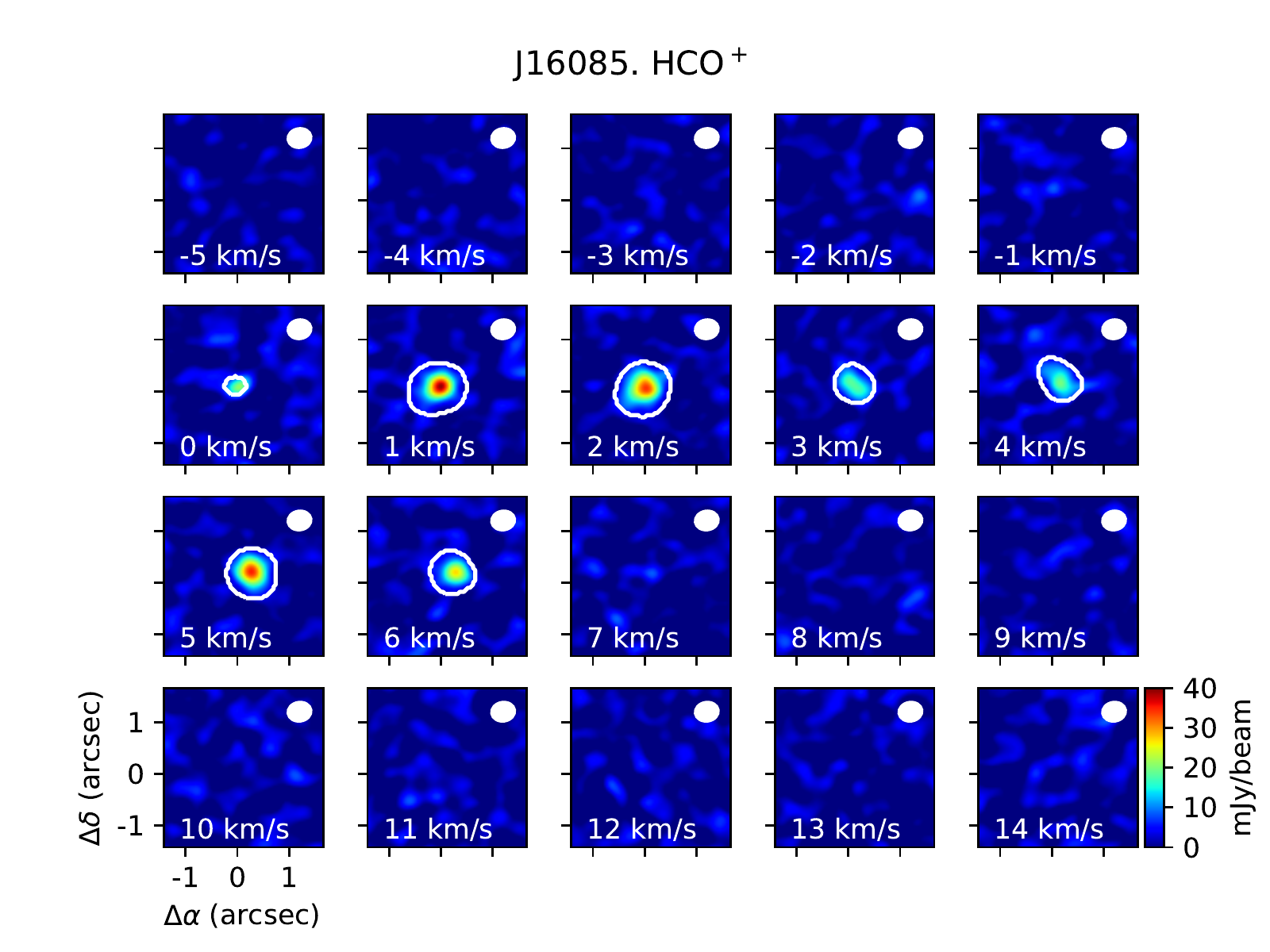}
  \includegraphics[width=\linewidth]{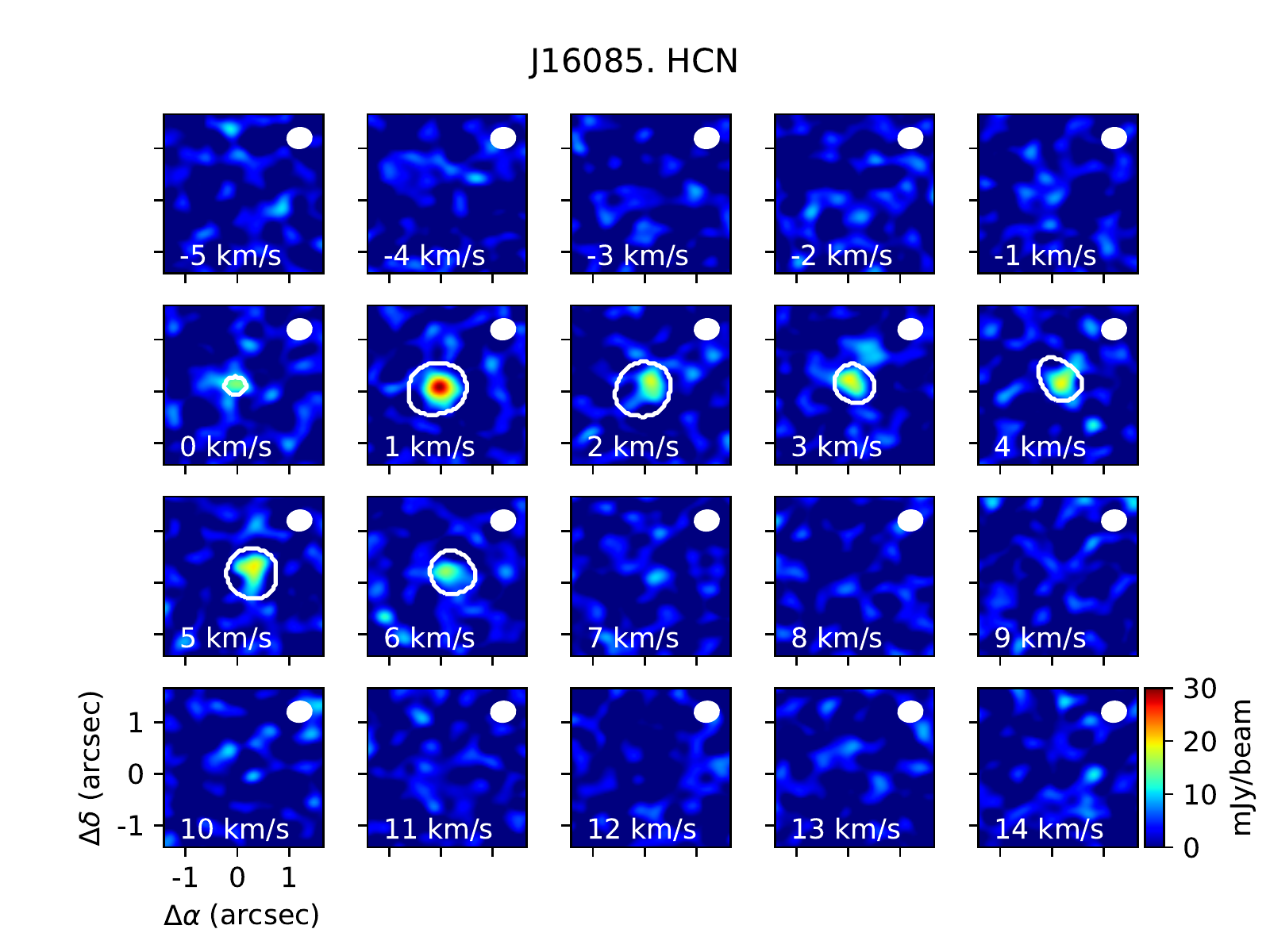}
  \includegraphics[width=\linewidth]{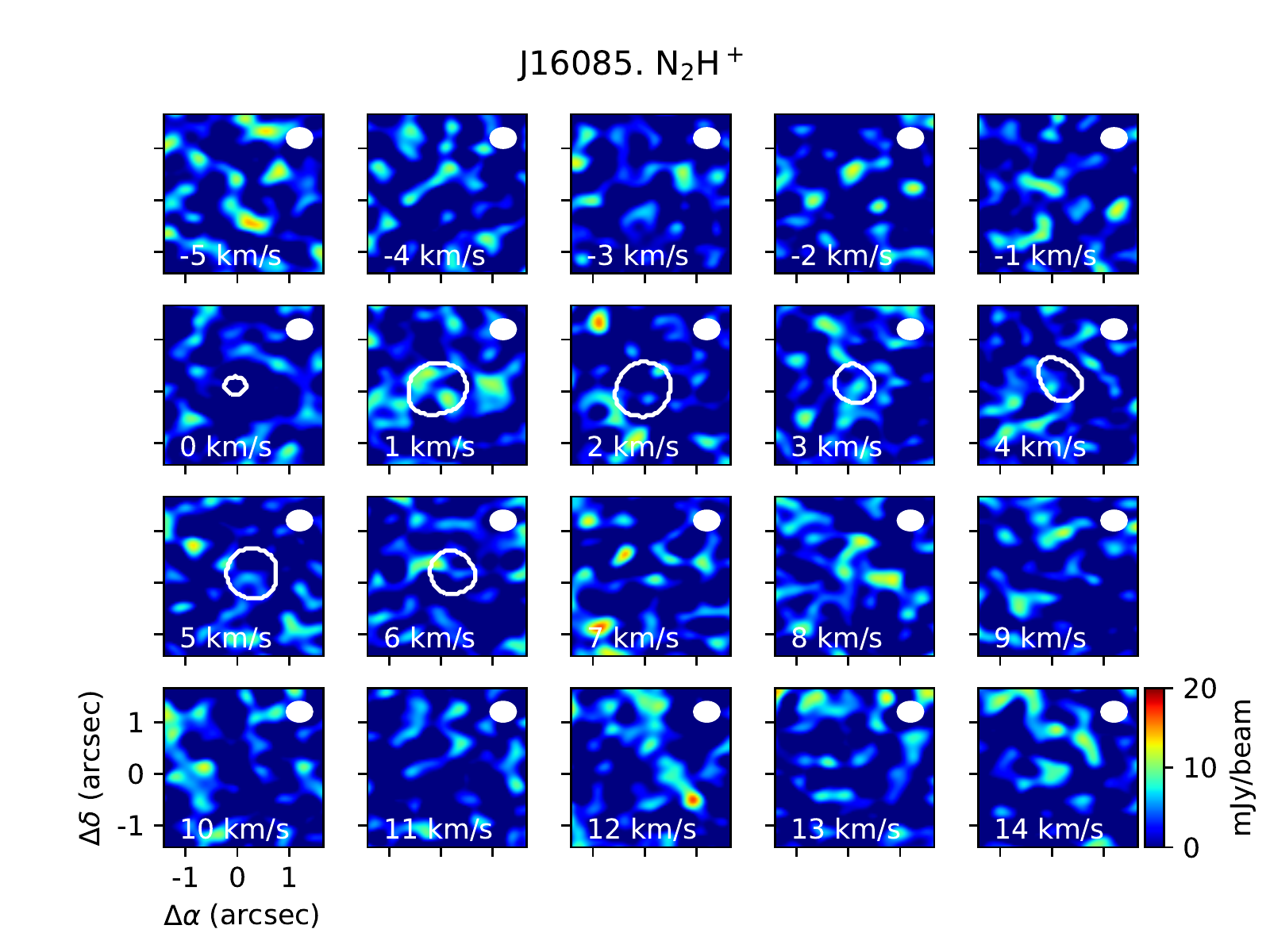}
  \captionsetup{width=0.95\linewidth}
  \caption{Channel maps of HCO$^+$, HCN, and N$_2$H$^+$ line emission for 2MASSJ16085324-3914401. Contours outline the mask generated based on the HCO$^+$ emission. Tick marks indicate 1\arcsec. The beam is shown in the upper right. }
  \label{fig:maps5}
\end{minipage}%
\begin{minipage}{.49\textwidth}
  \includegraphics[width=\linewidth]{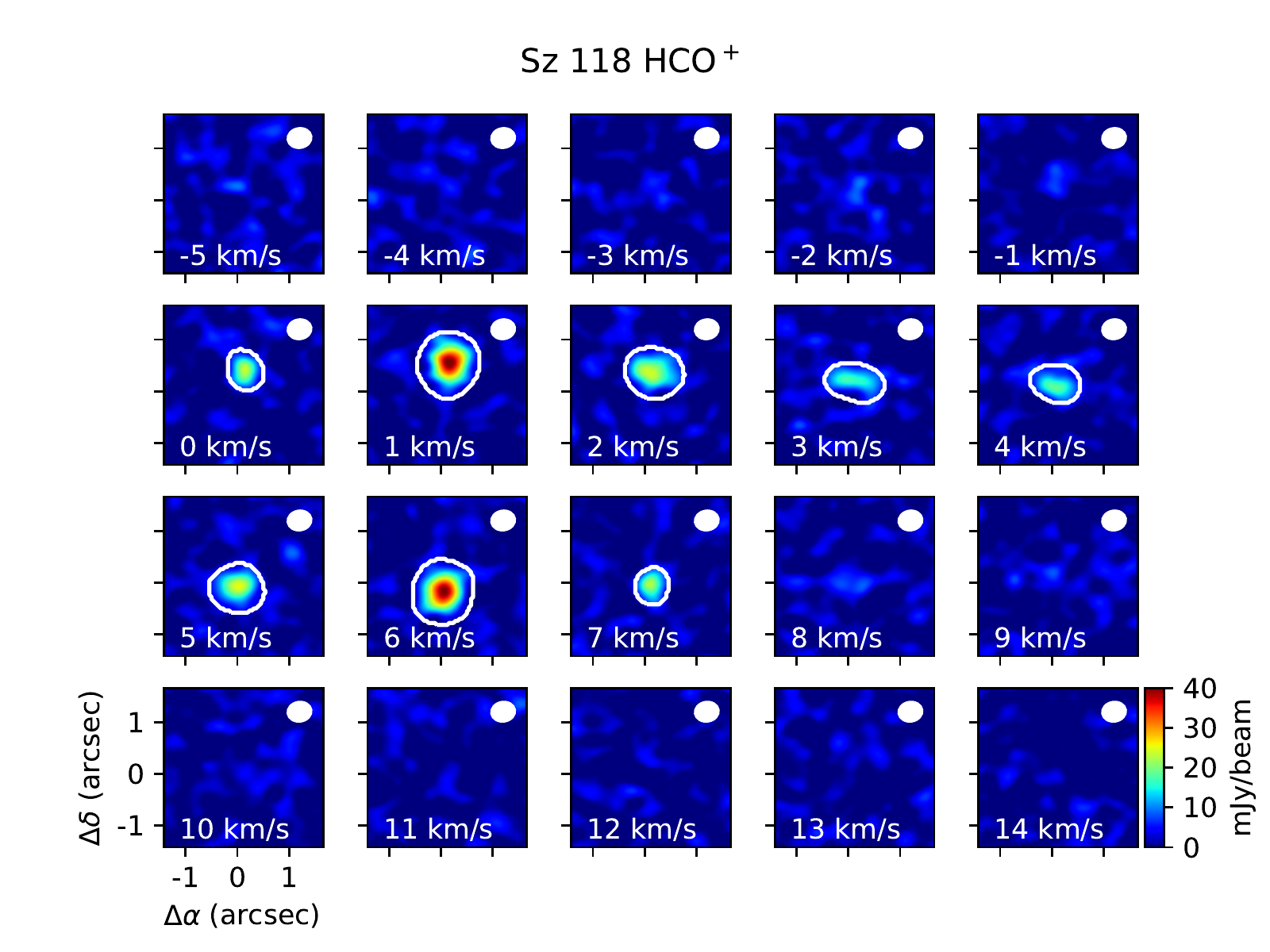}
  \includegraphics[width=\linewidth]{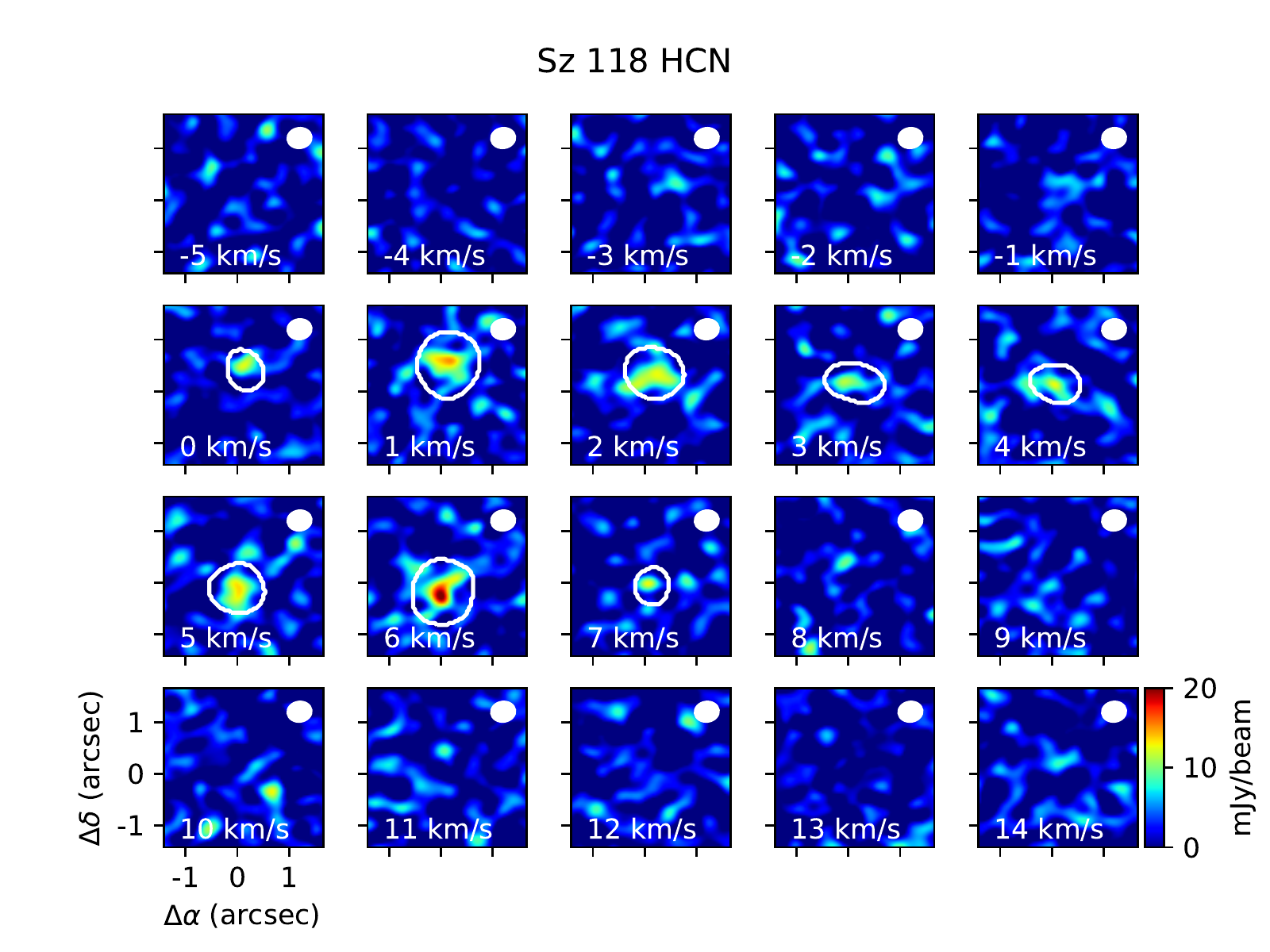}
  \includegraphics[width=\linewidth]{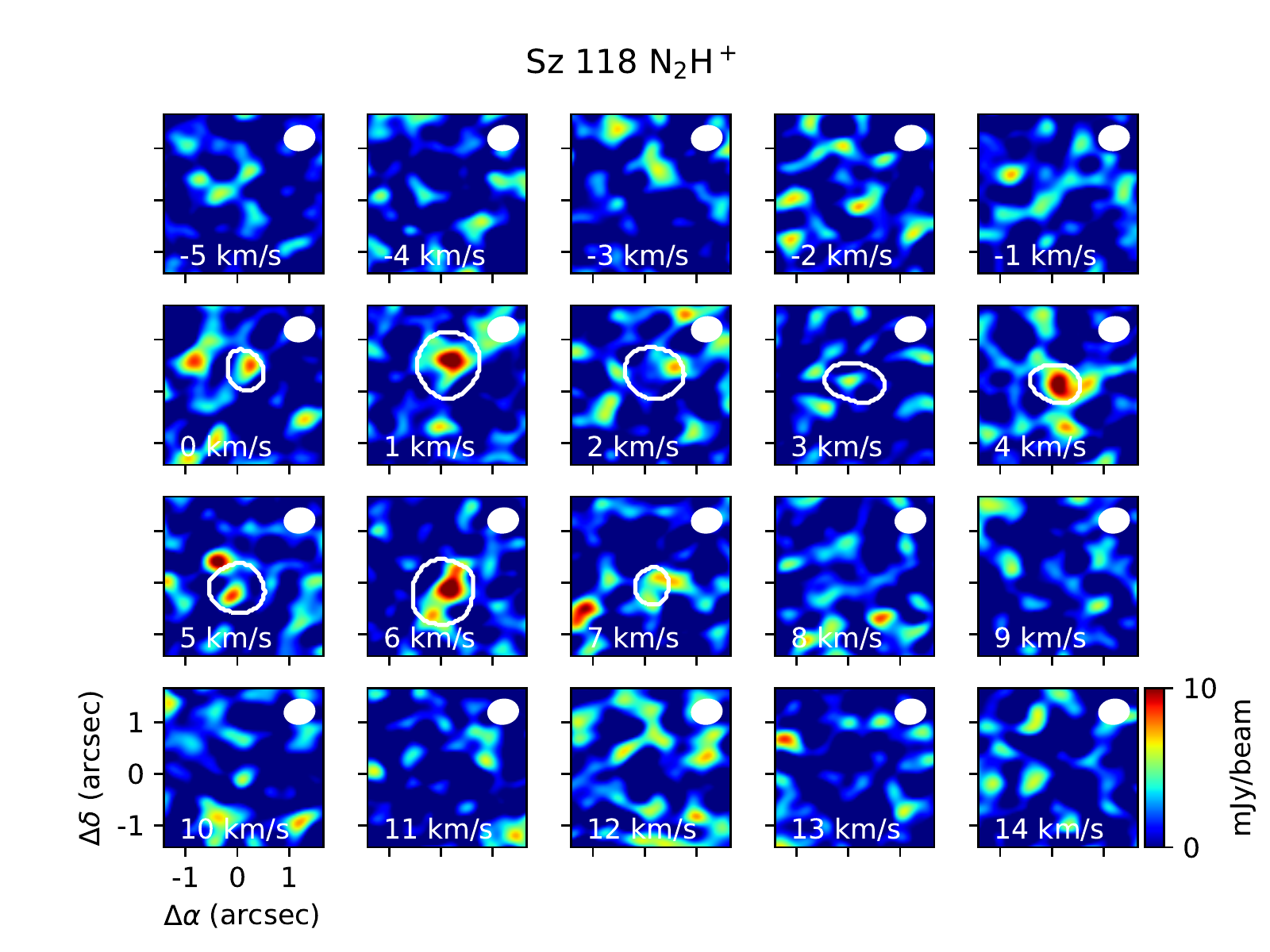}
  \captionsetup{width=0.95\linewidth}
  \caption{Channel maps of HCO$^+$, HCN, and N$_2$H$^+$ line emission for Sz~118. Contours outline the mask generated based on the HCO$^+$ emission. Tick marks indicate 1\arcsec. The beam is shown in the upper right.
}
\label{fig:maps6}
\end{minipage}
\end{figure}

\begin{figure}
\begin{minipage}{.49\textwidth}
  \includegraphics[width=\linewidth]{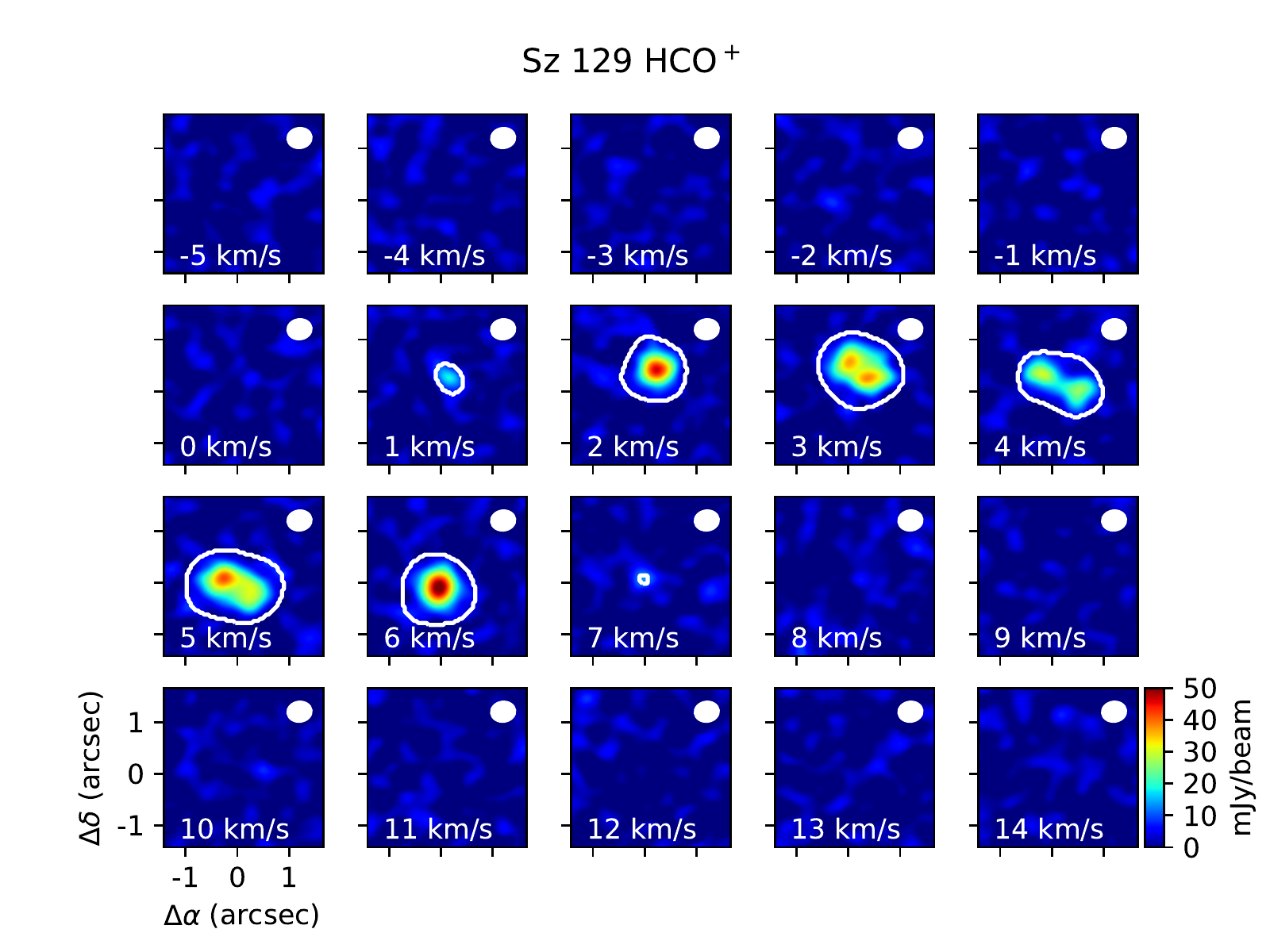}
  \includegraphics[width=\linewidth]{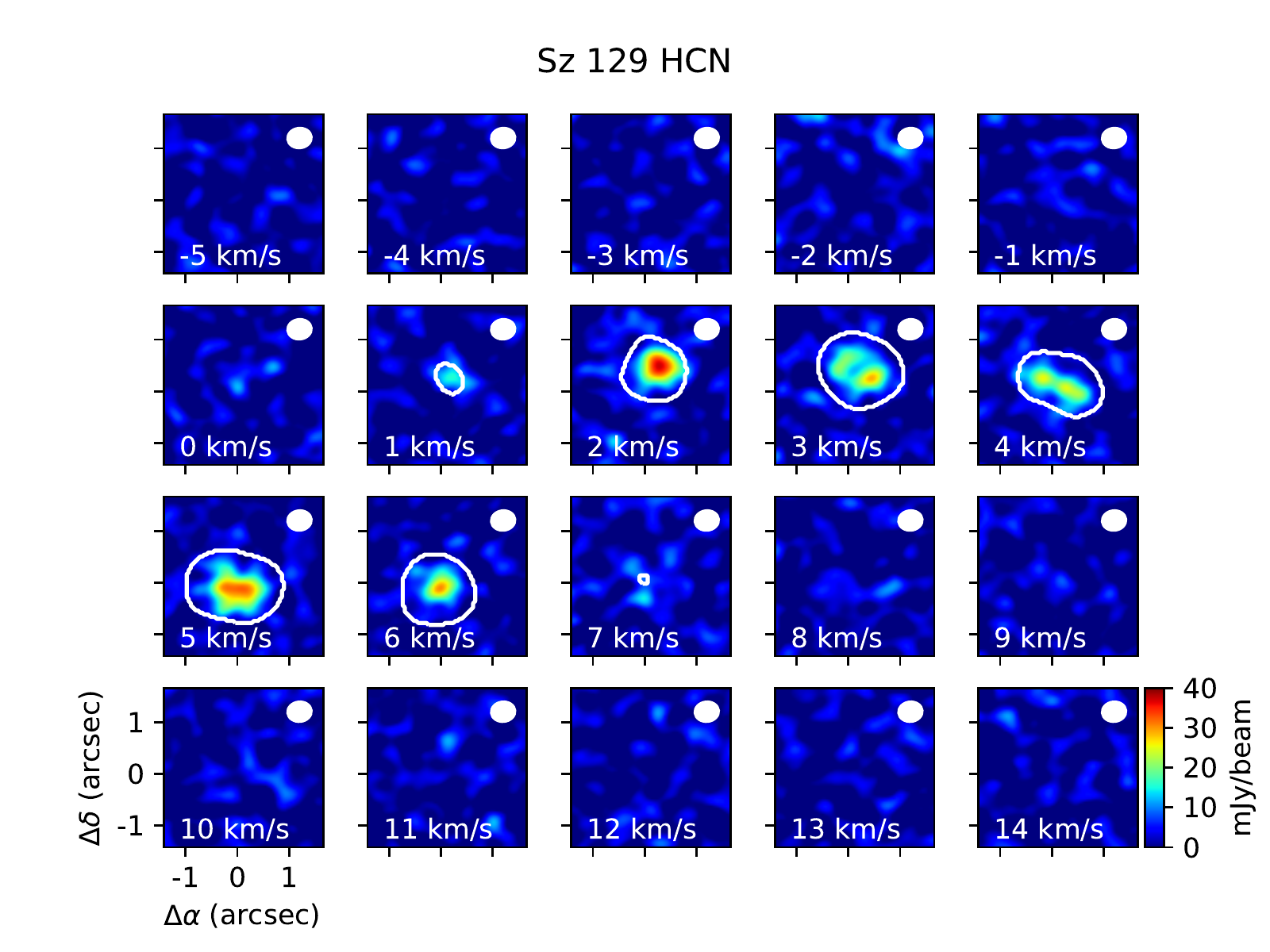}
  \includegraphics[width=\linewidth]{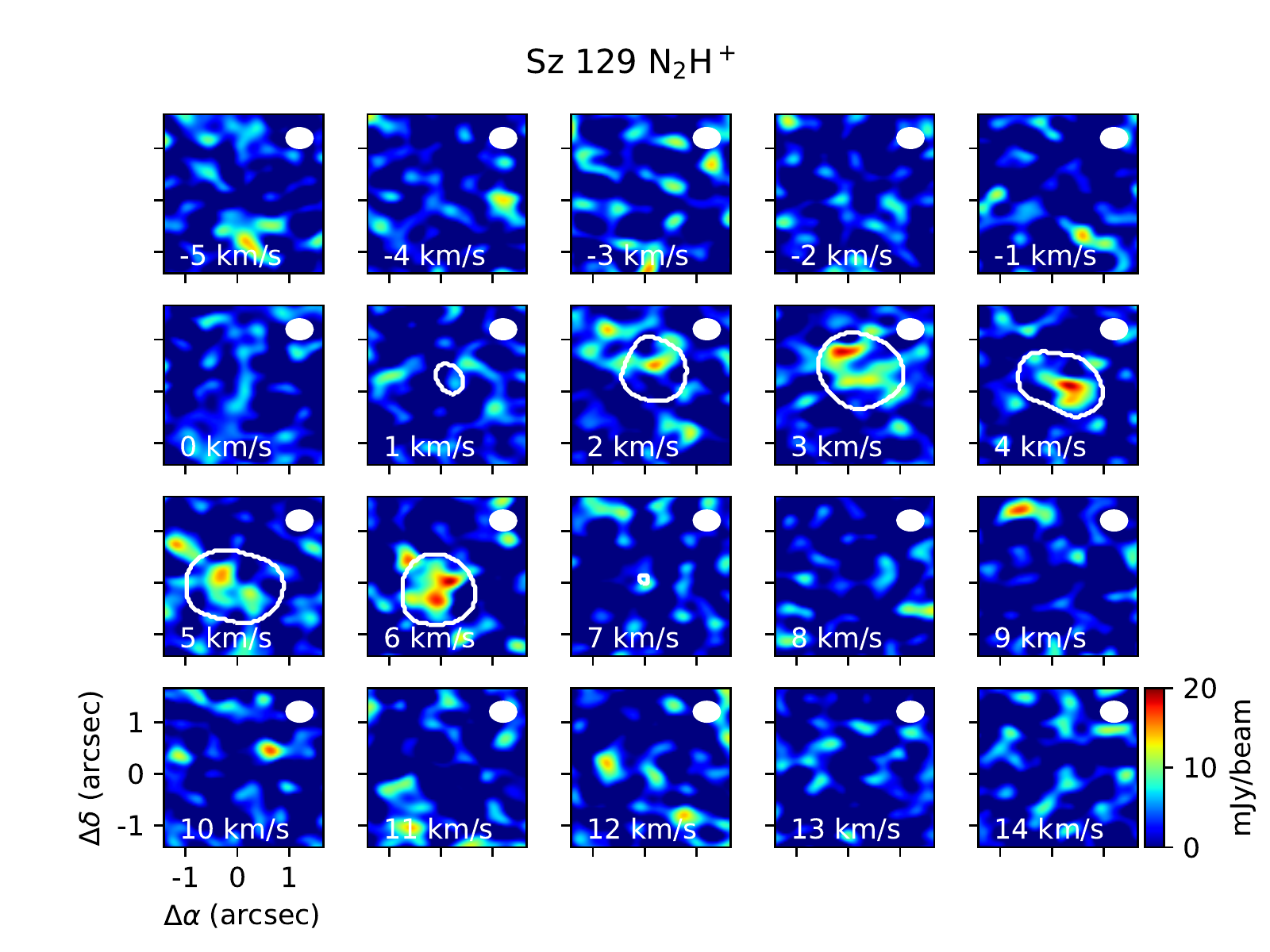}
  \captionsetup{width=0.95\linewidth}
  \caption{Channel maps of HCO$^+$, HCN, and N$_2$H$^+$ line emission for Sz~129. Contours outline the mask generated based on the HCO$^+$ emission. Tick marks indicate 1\arcsec. The beam is shown in the upper right.
}
\label{fig:maps7A}
\end{minipage}
\end{figure}

\appendix
Figure~\ref{fig:env} shows the physical structure of our fiducial disk model. The channel maps for all seven sources are provided in Figures~\ref{fig:maps1}--\ref{fig:maps7A}.

\end{document}